\documentclass[twocolumn]{aastex631}

\usepackage{appendix}

\definecolor{darkgoldenrod}{rgb}{0.72, 0.53, 0.04}

\usepackage{xcolor}

\begin{document}

\title{\texttt{McFACTS} I: Testing the LVK AGN channel with Monte Carlo For AGN Channel Testing \& Simulation (\texttt{McFACTS})}

\author[0000-0002-0786-7307]{Barry McKernan}
\affiliation{Center for Computational Astrophysics, Flatiron Institute, 
162 5th Ave, New York, NY 10010, USA}
\affiliation{Department of Astrophysics, American Museum of Natural History, New York, NY 10024, USA}
\affiliation{Department of Science, BMCC, City University of New York, New York, NY 10007, USA}
\affiliation{Graduate Center, City University of New York, 365 5th Avenue, New York, NY 10016, USA}

\author[0000-0002-5956-851X]{K.E. Saavik Ford}
\affiliation{Center for Computational Astrophysics, Flatiron Institute, 
162 5th Ave, New York, NY 10010, USA}
\affiliation{Department of Astrophysics, American Museum of Natural History, New York, NY 10024, USA}
\affiliation{Department of Science, BMCC, City University of New York, New York, NY 10007, USA}
\affiliation{Graduate Center, City University of New York, 365 5th Avenue, New York, NY 10016, USA}

\author[0000-0001-7163-8712]{Harrison E. Cook}
\affiliation{New Mexico State University, Department of Astronomy, PO Box 30001 MSC 4500, Las Cruces, NM 88001, USA}

\author[0000-0001-7099-765X]{Vera Delfavero}
\affiliation{Gravitational Astrophysics Laboratory, NASA Goddard Space Flight Center, Greenbelt, MD 20771, USA}

\author[0009-0008-5622-6857]{Emily McPike}
\affiliation{Department of Astrophysics, American Museum of Natural History, New York, NY 10024, USA}
\affiliation{Graduate Center, City University of New York, 365 5th Avenue, New York, NY 10016, USA}

\author[0000-0003-2430-9515]{Kaila Nathaniel}
\affiliation{Center for Computational Relativity and Gravitation, Rochester Institute of Technology, Rochester, New York 14623, USA}

\author[0000-0003-0738-8186]{Jake Postiglione}
\affiliation{Department of Astrophysics, American Museum of Natural History, New York, NY 10024, USA}
\affiliation{Graduate Center, City University of New York, 365 5th Avenue, New York, NY 10016, USA}

\author[0009-0005-5038-3171]{Shawn Ray}
\affiliation{Department of Astrophysics, American Museum of Natural History, New York, NY 10024, USA}
\affiliation{Graduate Center, City University of New York, 365 5th Avenue, New York, NY 10016, USA}

\author[0000-0001-5832-8517]{Richard O'Shaughnessy}
\affiliation{Center for Computational Relativity and Gravitation, Rochester Institute of Technology, Rochester, New York 14623, USA}



\begin{abstract}
Active galactic nuclei (AGN) are a promising source of the binary black hole (BBH) mergers observed in gravitational waves with LIGO-Virgo-Kagra (LVK). Constraining the AGN channel allows us to limit AGN parameter space (disk density, size, average lifetime) and nuclear star cluster (NSC) parameter space. Constraints on AGN and NSCs have implications for $\Lambda$CDM models of AGN feedback and models of AGN-driven SMBH merger and growth. Here we present several qualitative studies of the AGN channel using new public, open-source, fast, reproducible code \texttt{McFACTS}\footnote{https://github.com/mcfacts/mcfacts}:Monte Carlo for AGN channel Testing \& Simulation. We demonstrate several important features for testing the AGN channel, including: i) growth to large mass IMBH is helped by the presence of migration traps or swamps, ii) flat BH initial mass functions highlight hierarchical merger features in the mass spectrum, iii) the ($q,\chi_{\rm eff}$) anti-correlation is a strong test of the bias to prograde mergers in the AGN channel, iv) spheroid encounters can drive a fraction of mergers with high in-plane spin components ($\chi_{\rm p}$), v) a high rate of extreme mass ratio inspirals (EMRIs) are driven by an initial population of embedded retrograde BH, vi) Both LVK and LISA are powerful probes of models of AGN disks and their embedded populations.

\end{abstract}



\section{Introduction} \label{sec:intro}
Active Galactic Nuclei (AGN) are a promising source of binary black hole (BBH) mergers observed in gravitational waves (GW) \citep{McK14,Bartos17,Stone17,ArcaSedda23,Ford25}. A fraction of the stellar mass black holes (BH) and other objects in a nuclear star cluster (NSC) must have orbits coincident with the plane of AGN disks. Another fraction of NSC orbiters can be captured by the AGN disk during its lifetime \citep{Artymowicz93,MacLeod20,Fabj20,Nasim22,WZL24}. 

BH embedded in an AGN disk can migrate and encounter each other, potentially forming BBH \citep[e.g.][]{Stan23,Rowan23,Rozner23,Qian24,DodiciTremaine24}. Gas hardening and/or dynamical encounters can drive rapid BBH mergers \citep{Bellovary16,Leigh18,Yang19,McK20a,Secunda21,Rixin22,Jairu23}. Simulations of AGN mergers (both N-body and Monte-Carlo) yield commonly expected results from this channel. First, intermediate mass black hole (IMBH) formation is more efficient than in any other LVK channel \citep{Hiromichi20,Secunda21}, since high recoil kicks at merger do not permit escape from inner AGN disks where Keplerian velocity is $\mathcal{O}(10^{4}\, {\rm km/s}\, (a/10^{3}\,r_{g}))$ with semi-major axis $a$ in units of $r_{g}=GM_{\rm SMBH}/c^{2}$ the SMBH gravitational radius \citep[see also][]{Gerosa19,Ford22,Xue25}. Second, migrations within disks promote more asymmetric mass mergers then expected from channels biased towards equal mass mergers \citep{McK20b}. Third, a bias away from a random distribution of spins due to symmetry-breaking from gas accretion and torquing by the AGN disk \citep{Hiromichi20spin,AlexD24}. Fourth, residual eccentricity in the LVK band may be more common among BBH in AGN than in any other channel \citep[e.g.][]{Samsing22,Calcino23}. Fifth, in-plane spin components may be more common among BBH in AGN than in any other channel \citep[e.g.][]{Hiromichi20spin,McKF24}. Sixth, while BH-NS mergers can occur commonly in AGN disks, NS-NS mergers are sub-dominant since migration is mass dependent and only pre-existing hard NS-NS binaries can merge \citep{McK20b}. 

Unlike any other LVK channel, EM counterparts \emph{must} always occur for AGN BBH mergers, surrounded as they are by magnetized gas \citep{Bartos17,McK19,Hiromichi24}. The detectability of EM counterparts depends on whether they can: (i) emerge from the optically thick disk mid-plane on the observer's side (not observable in Type II AGN or in half of Type I AGN), (ii) be luminous enough to be detectable versus intrinsic AGN luminosity and (iii) be distinguishable from potential false positives \citep[e.g.][]{Graham20,Graham23,Cabrera24,KimElias24}. Constraints can be placed on the AGN fractional contribution to the LVK rate by testing the overlap between flaring AGN and LVK GW event volumes \citep[e.g.][]{Bartos17stats,Veronesi23}.

Clues from LVK results so far include: a relatively high rate of IMBH formation (roughly $10\%$ of events in O3), a bias towards positive $\chi_{\rm eff}$, evidence for multiple events with in-plane spin components ($\chi_{\rm p} >0$), possible hints of eccentricity \citep{Isobel22} and an intriguing anti-correlation in ($\rm{q},\chi_{\rm eff}$) \citep{Callister21} that could arise in a relatively straightforward manner in AGN disks \citep{qX22,Santini23}.

Constraints on the fraction ($f_{\rm AGN}$) of LVK BBH mergers that come from AGN as well as properties of those mergers in turn allow us to constrain average AGN properties, like disk size, disk density and disk duration \citep{McK18,Gayathri21,Vajpeyi22,Gayathri23}. Since we cannot directly resolve AGN disks, this allows us to constrain the role of AGN in galactic feedback in $\Lambda$CDM and in bringing merging SMBH together (the ``final pc'' problem) \citep{Ford25}.

BH embedded in an AGN disk can also lead to a population of Extreme Mass Ratio Inspiral (EMRI) systems, or stellar origin black holes merging with the central SMBH. Predictions for the EMRI population are poorly constrained, but will be measureable with LISA  \citep{Amaro_Seoane_2007,Oliver2024}.

Population synthesis (pop-synth) codes are valuable Astrophysics tools for studying a population based on user-chosen model assumptions. Such codes are particularly useful in testing poorly constrained conditions, which would be required to reproduce overall population properties. Excellent and standard-setting modern examples of such codes include: \texttt{COSMIC} \citep{COSMIC20}, \texttt{COMPAS} \citep{Riley2022}, \texttt{CMC} \citep{CMC22} and \texttt{cogsworth} \citep{cogsworth24}. The AGN channel for LVK/LISA is remarkably uncertain \citep{McK18} and therefore can usefully be tested by a pop synth code. Pop synth codes have been developed for the AGN channel by several groups (including us), but typically focus on specific features of the AGN channel, e.g. mergers at migration traps \citep{Yang19,Vaccaro24}, or are private \citep{McK20a,Hiromichi20}.

Here we demonstrate the new, public, open source, reproducible code $\texttt{McFACTS}$ which allows for the rapid testing of BBH merger properties across AGN and NSC parameter space. The code as released (\texttt{v.0.3.0}) takes a broad brushstroke approach to the AGN channel, making multiple simplifying assumptions and is missing detailed physical models across parameter space. However, \texttt{McFACTS} is intended to be a living, developing code (see \S\ref{sec:future} below for planned near-future additions). Here we illustrate the code by demonstrating some qualitative results and dependencies, and show how AGN channel results can be tested against LVK observations. We reserve detailed studies of particular features of the AGN channel using \texttt{McFACTS} to companion papers, including: testing the observed $(q,\chi_{\rm eff})$ anti-correlation \citep{Cook24}, measuring the detectable population of BBH mergers as a function of Galaxy mass in a model Universe \citep{Delfavero24}, testing the role of stars in the AGN channel ({Nathaniel et al., 2025, in prep.}), testing the population of EMRI events expected ({Ford et al. 2025, in prep.}) and testing models of gas hardening ({Postiglione et al. 2025, in prep.}). We anticipate that \texttt{McFACTS} will be useful in testing correlations between AGN parameters and GW source populations, allowing refinement of work by \citep{Bartos17stats,Veronesi23}.

\section{The Code}
\label{sec:code}

Here we outline the released \textbf{\texttt{v.0.3.0}} of the code including overall structure, initial conditions, physical mechanisms and approximations, and standard as well as optional outputs. Full documentation can be found at \texttt{https://github.com/mcfacts}. The code is written in \texttt{Python} for ease of readability and use. \texttt{McFACTS} \textbf{\texttt{v.0.3.0}} is fast, with a grid of 100 AGN galaxy realizations running in $550-600\, {\rm s}$ on an Apple M1 Pro (Model Identifier: MacBookPro18,1). This test was carried out with default settings, for $100$ timesteps of duration $10^4 \,\mathrm{yr}$. See \citep{Delfavero24} for performance scaling with SMBH mass.

\begin{figure*}   \includegraphics{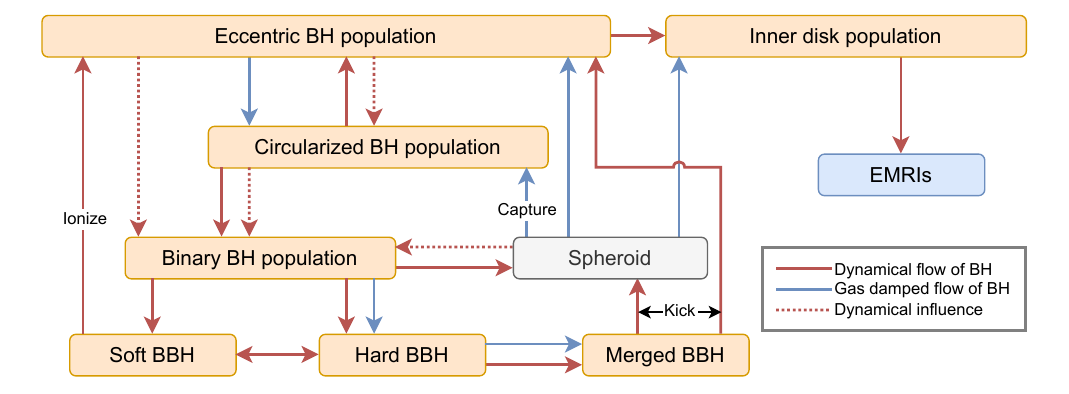}
    \caption{Flow diagram indicating how BH are categorized in \texttt{McFACTS} and how BH are exchanged between these categories. Solid arrows indicate a flow of objects from one category to another and dotted arrows indicate an influence of one category on another. Red denotes dynamics, blue denotes gas. Objects in orange (grey) boxes are in the disk (spheroid). EMRIs are in the blue box. For example, there is a flow from the initially eccentric BH population to the circularized BH population due to gas damping (blue arrow), but the eccentric BH population also has a dynamical influence on the circularized BH population (red dotted line) leading to a reverse flow (red solid line). 
    }
    \label{fig:overview}
\end{figure*}

Fig.~\ref{fig:overview} gives a basic overview of the flow of BH through categories in \texttt{McFACTS}, starting from an initial eccentric BH population of singleton BH in the AGN disk (top left). Solid arrows indicate a flow from one category to another and dotted arrows indicate influence from one category on another. For example, there is a flow from the initial eccentric BH population into the circularized BH population via gas damping (blue solid arrow), but the eccentric BH population also dynamically interacts with the circularized population (dotted red arrow) and can excite it to reverse the flow (solid red arrow) . Binary BH are formed from among the circularized population (connecting solid red arrow). BBH, once formed can dynamically interact with both the eccentric and circularized single BH and the spheroid population (the three dotted red arrows onto Binary BH in Fig.~\ref{fig:overview}). BBH can flow to softer BBH and ionize, via dynamics (solid red arrows) returning their components back to the eccentric disk BH population, but BBH can also harden via gas and dynamics (blue and red arrows) and ultimately merge, whereupon they either flow back into the eccentric BH population or into the spheroid (from where they can be re-captured generally circularized). A small sub-population of the initial eccentric population and the spheroid population can flow into the inner disk, where some BH can decay via GW emission and become EMRIs (blue box).

Objects in their disks have their physical parameters manipulated in a consistent algorithmic manner via the \verb|AGNObject| class. This class will be extended from just BH to stars, neutron stars, and white dwarfs in future updates to the code. Each \verb|AGNObject| is given a unique ID number, stored in an instance of the \verb|AGNFilingCabinet| class, along with other parameters such as the object's mass and location in the AGN disk. This allows the code to iterate over the entire population of disk objects when calculating e.g. interactions and binary formation, rather than separating and evaluating objects by type for each iteration.

\subsection{Initial conditions and other inputs}
There are three basic inputs for any simulated galactic nucleus: the SMBH, the NSC, and the AGN disk. The initial conditions for each of these is set via an input file, for which a default file is provided, or constructed by the user (see Table~\ref{tab:default_model} for \texttt{p1$\_$default$\_$model.ini}). Primary input parameters are the SMBH mass ($M_{\rm SMBH}$), the NSC mass ($M_{\rm NSC}$) and the AGN disk model type---users may choose from either a \cite{SG03} or \cite{TQM05} type disk, as implemented in the \texttt{pAGN} package \citep{pAGN24} (and for which the user must also specify the disk $\alpha$ parameter, mass accretion rate, $\dot{m}$, and radiative efficiency $\eta$), or may supply their own disk model if they can provide tabulated surface density ($\Sigma$), aspect ratio ($h/r$), and opacity ($\tau$) profiles as a function of disk radius (subject to the requirement that a user-supplied model must be interpolatable over the range of disk radii under consideration, see also below).
A simplified implementation of both the \cite{SG03} and \cite{TQM05} disk models which assumes constant opacity is also provided in \texttt{McFACTS} along with the pAGN implementation.

\begin{deluxetable}
{r|cc}
\label{tab:default_model}
\tabletypesize{\scriptsize}
\tablewidth{0pt} 
\tablecaption{Default input parameters for the SMBH, AGN disk and NSC as well as physics choices in \texttt{p1$\_$model$\_$default.ini}.}
\tablehead{
\colhead{Parameter} & \colhead{Default value} &\colhead{Units}\\
}
\startdata
\texttt{smbh$\_$mass} & $10^{8}$ & $M_{\odot}$ \\
\hline
\texttt{disk$\_$model$\_$name} & \texttt{sirko$\_$goodman} & \\
\texttt{flag$\_$use$\_$pagn} & 1 & \\
\texttt{disk$\_$radius$\_$trap} & 700 & $\rm{r}_{g,\rm{SMBH}}$ \\
\texttt{disk$\_$radius$\_$outer} & $5\times 10^{4}$ & $\rm{r}_{g,\rm{SMBH}}$ \\
\texttt{disk$\_$radius$\_$max$\_$pc} & 0 & \\
\texttt{disk$\_$alpha$\_$viscosity} & 0.01 &  
\\
\texttt{inner$\_$disk$\_$outer$\_$radius} & $50$ & $\rm{r}_{g,\rm{SMBH}}$ \\
\texttt{disk$\_$inner$\_$stable$\_$circ$\_$orb} & $6$ & $\rm{r}_{g,\rm{SMBH}}$ \\
\texttt{disk$\_$aspect$\_$ratio$\_$avg} & $0.03$ &  \\
\hline
\texttt{nsc$\_$radius$\_$outer} & 5 & pc\\
\texttt{nsc$\_$radius$\_$crit} & 0.25 & pc\\
\texttt{nsc$\_$mass} & $3 \times 10^{7}$ & $M_{\odot}$ \\
\texttt{nsc$\_$ratio$\_$bh$\_$num$\_$star$\_$num} & $10^{-3}$ &  \\
\texttt{nsc$\_$ratio$\_$bh$\_$mass$\_$star$\_$mass} & 10 &  \\
\texttt{nsc$\_$density$\_$index$\_$inner} & 1.75 & \\
\texttt{nsc$\_$density$\_$index$\_$outer} & 2.5 & \\
\texttt{nsc$\_$spheroid$\_$normalization} & 1 &  \\
\texttt{nsc$\_$imf$\_$bh$\_$mode} & 10 & $M_{\odot}$ \\
\texttt{nsc$\_$imf$\_$bh$\_$max} & 40 & $M_{\odot}$ \\
\texttt{nsc$\_$imf$\_$powerlaw$\_$index} & 2 & \\
\texttt{nsc$\_$bh$\_$spin$\_$dist$\_$mu} & 0 & \\
\texttt{nsc$\_$bh$\_$spin$\_$dist$\_$sigma} & 0.1 & \\
\texttt{nsc$\_$imf$\_$powerlaw$\_$index} & 2 & \\
\hline
\texttt{disk$\_$bh$\_$torque$\_$condition} & 0.1 & \\
\texttt{disk$\_$bh$\_$eddington$\_$ratio} & 1.0 & \\
\texttt{disk$\_$bh$\_$orb$\_$ecc$\_$max$\_$init} & 0.3 & \\
\texttt{disk$\_$radius$\_$capture$\_$outer} & $10^{3}$ & $\rm{r}_{g,\rm{SMBH}}$ \\
\texttt{disk$\_$bh$\_$pro$\_$orb$\_$ecc$\_$crit} & 0.01 & \\
\texttt{mass$\_$pile$\_$up} & 35 & $M_{\odot}$\\
\texttt{timestep$\_$duration$\_$yr} & $10^{4}$ & yr\\
\texttt{timestep$\_$num} & 100 & \\
\texttt{capture$\_$time$\_$yr} & $10^{5}$ & yr\\
\texttt{galaxy$\_$num} & 100 & \\
\texttt{delta$\_$energy$\_$strong} & 0.1 & \\
\texttt{fraction$\_$bin$\_$retro} & 0.0 & \\
\texttt{flag$\_$thermal$\_$feedback} & 1 & \\
\texttt{flag$\_$orb$\_$ecc$\_$damping} & 1 & \\
\texttt{flag$\_$dynamic$\_$enc} & 1 & \\
\texttt{flag$\_$torque$\_$prescription} & \texttt{paardekooper} & \\
\texttt{mean$\_$harden$\_$energy$\_$delta} & 0.9 & \\
\texttt{var$\_$harden$\_$energy$\_$delta} & 0.025 & \\
\hline
\enddata
\tablecomments{We do not include initialization of star parameters since stars are turned off by default in \texttt{v.0.3.0}.}
\end{deluxetable}

We do not currently consider the spin of the SMBH, so there are no input parameters related to it. However the NSC and the AGN disk have a number of important additional parameters that the user can specify (or use the defaults provided). Given $M_{\rm NSC}$ and a disk model, we determine (purely based on geometry) the number of stellar mass black holes with orbits fully embedded in the AGN disk at the time the AGN disk initially arrives in the nucleus (here we make the simplifying assumption that the disk does not evolve through the simulation, appearing instantaneously at time zero, subsequently vanishing instantaneously at time $\tau_{\rm AGN}$, the lifetime of the AGN disk, also specified in the input file). The remainder of the NSC parameters are for distribution functions from which we select the actual objects included in a given simulation for each galaxy. The user can specify a random number seed (\texttt{SEED}) in the \texttt{Makefile} or take the default (in either case the value is recorded for reproducibility). If the user is simulating multiple galaxies with the same input parameters (the usual mode of operation, in order to build up statistics), each new galaxy will (by default) iterate the original random seed by one, ensuring an entire suite of simulations can be easily reproduced from a starting \texttt{SEED} and \texttt{.ini} file.

The \texttt{SEED} is used to generate initially embedded stellar black hole (BH) masses, orbital semi-major axes ($a$), eccentricities ($e$), inclinations ($i$), arguments of periapse ($\omega$), spin magnitude ($\chi$), and spin angle ($\theta$) from initial distributions set by the \texttt{.ini} file. BH masses are drawn from an initial mass function (IMF), assumed to be a Pareto function with a mode mass and powerlaw slope, each of which the user can specify in the \texttt{.ini} file. An additional mass feature can be added to the IMF by piling up masses drawn from the powerlaw IMF beyond some maximum mass in a user specified Gaussian distribution. In this way we can emulate a possible pair instability supernova (PISN) bump in the BH IMF, as physically expected from models of stellar evolution \citep[e.g.][]{Renzo24}, as well as a mass pile-up in the LVK observations \citep{massbump23}. BH number density is set by the NSC model and default is a broken power-law distribution consistent with NSC models \citep{Generozov18,Neumayer20}. The number of BH geometrically coincident with the disk between disk inner and outer radii is then calculated and BH orbital semi-major axes are chosen from a (default) broken power law distribution consistent with the NSC radial density distribution, although the user can also draw from a uniform distribution. Orbital eccentricities are drawn uniformly between zero and a (user specified) maximum initial eccentricity by default but the user can also specify a thermal initial distribution (likely the limiting case where a population has fully relaxed between AGN episodes). Orbital inclinations of embedded BH are selected uniformly between $\pm h/r$, the disk aspect ratio, where $h$ is the disk height from the midplane at radius $r$. 

Next we assign an orbital angular momentum flag of $\pm 1$ to specify either prograde or retrograde orbits with respect to the disk gas, allowing us to apply appropriate physics to the different objects (and implicitly setting the inclinations to be either $i=0\pm h/r$ (prograde) or $i=\pi \pm h/r$ (retrograde)) where orbiters with $i=0^{\circ}$ lie prograde in the disk midplane. The argument of periapse for each object is currently chosen to be either $0$ or $\pi/2$ (see \S\ref{sec:singlebh} below). We recognize this is a limitation of the code as the timescales for capture depend strongly on the argument of periapse and differ by orders of magnitude between these two values. This decision will overestimate the capture efficiency at early times, and underestimate the capture efficiency at late times, but over the course of a typical run, these behaviors should roughly average out in the population statistics. \footnote{In reality a random distribution across a wider range of values would be more appropriate; however, the form of the evolution of an orbiter is also quite disparate between these two values. Interpolating between them, or, for even greater precision, passing the orbital parameters to an N-body code, would add substantial computational expense, so we currently leave the two extremal behaviors as the only options.  We plan to add \textit{slightly} finer divisions in a future release}. BH spin magnitude (dimensionless spin parameter) is chosen from a Gaussian distribution defined by a user specified mean and standard deviation. BH spin angles ($\theta$, w.r.t. the AGN disk orbital angular momentum) are chosen uniformly between $[0,\pi/2]\,([\pi/2,\pi])$(radians) for BH with initial positive (negative) spin parameters, where $\theta=0(\pi)$ is fully aligned(anti-aligned) with the AGN disk gas. If the disk model is expected to contain a migration trap, the user must specify the radius. A disk outer radius must also be specified, with the default disk inner radius assumed to be the innermost stable circular orbit (ISCO) for a non-spinning (Schwarzschild) SMBH. 

Finally, the user must specify the duration of a timestep in years and the number of timesteps per galaxy, the product of which specifies the total disk lifetime ($\tau_{\rm AGN}$, plausibly $\sim 0.1-10{\rm Myr}$). A default timestep duration of $10^{4}$ yr provides a reasonable compromise between computational speed and the resolution of the physical processes being modeled. Users who wish to see finer detail on particular physical processes or individual events may choose smaller timesteps. 

\subsection{Physical processes modelled}

\subsubsection{Single BH in the disk}
\label{sec:singlebh}
For every galaxy, we compute the number of black holes which should be coincident with the AGN disk, based on the user specified parameters for the disk and NSC, based on geometry. We then draw the requisite number of black holes from the distributions specified by the user and find, for their orbit around the SMBH, a semi-major axis, eccentricity, inclination and argument of periapse; we also find the mass, spin, and spin angle for each object. The initial binary fraction is zero, so all binaries in the simulation must be formed dynamically. We will allow for a user-specified initial binary fraction in future code updates. 

We then distinguish between prograde and retrograde orbiters, as they are subject to different processes. We also separate out the innermost disk orbiters (set by default to objects with semi-major axes $<50\,r_{g}$, adjustable by the user). We evolve orbits of inner disk objects\footnote{The user-defined inner disk is defined as the inner region of the disk within which purely GW decay could occur within a relatively long AGN lifetime. Default is $50r_{g}$, which corresponds to $\tau_{\rm GW} \sim 40$Myr around a default $M_{\rm SMBH}=10^{8}M_{\odot}$.} purely according to GR \citep[via the analytical formulae in][]{Peters64}, while outside of that boundary we evolve objects only due to gas torques and dynamical interactions between objects as described below. We will add gas effects to the evolution of the inner disk population in future code updates.

We evolve our collection of orbiters, subject to the various physics detailed below, and for each timestep we proceed as follows: For initially embedded prograde sBH we first compute gas processes which act on them. Objects which have eccentricities larger than the local disk aspect ratio ($e>h/r$) are not subject to migration torques. For each object over this eccentricity threshold we compute 
 characteristic orbital damping timescales for prograde orbits $t_{\rm damp}$ from \citet{Tanaka04}, re-written in \citet{McKF24}, as: 
\begin{equation}
    t_{\rm damp}=\frac{M_{\rm SMBH}^{2}(h/r)^{4}}{m_{\rm BH}\Sigma \,a^{2} \Omega}
\end{equation}
where $M_{\rm SMBH}$ is the supermassive black hole (SMBH) mass, $h$ is the disk scale height, $m_{\rm BH}$ is the embedded black hole mass, $\Sigma$ is the disk surface density, $a$ is now the orbital semi-major axis and $\Omega$ the Keplerian orbital frequency. We parameterize $t_{\rm damp}$ as:
\begin{eqnarray}
    t_{\rm damp} &\sim& 0.1\,{\rm Myr} \left(\frac{q}{10^{-7}}\right)^{-1}\left(\frac{(h/r)}{0.03}\right)^{4} \nonumber \\
    &\times& \left(\frac{\Sigma}{10^{5}\,{\rm kg \,m^{-2}}}\right)^{-1}\left(\frac{a}{10^{4}\,r_{g}}\right)^{-1/2}
\label{eq:damp}
\end{eqnarray}
where $q=m_{\rm BH}/M_{\rm SMBH}$ is the mass ratio of the embedded BH to the SMBH. 

For small initial orbital eccentricity ($e_{0}<2h$), we assume exponential decay \citep{Papaloizou00} 
\begin{equation}
    e(t)=e_{0}\,{\rm exp}(-t/t_{\rm damp})
\end{equation}
and for larger eccentricities ($e>2h$), we use the characteristic decay timescale \citep{Bitsch10,Horn12} 
\begin{equation}
    t_{e} \sim \frac{t_{\rm damp}}{0.78} \left[ 1 -0.14 (er/h)^{2} + 0.06 (er/h)^{3} \right].
\label{eq:te}
\end{equation}
In all cases, for the small initial inclination of any of the initially prograde objects we consider, the inclination damping timescale is shorter than the eccentricity damping timescale \citep[see e.g.][]{WZL24}, and so we do not separately or explicitly treat the damping. Instead we assume that the orbiter is brought to $i=0^{\circ}$ as long as $e<e_{\rm crit} \sim 0.01$(default).

For objects with sufficiently small eccentricities and inclinations, we expect them to migrate according to a Type I migration prescription\footnote{There are some objects in our simulations which may grow large enough to be subject to type 2 migration, but these are rare and we have not yet implemented type 2 migration in the code.}. We have included a user-set flag {\texttt{flag\_torque\_prescription} }which allows the user to choose between a 2-d torque prescription from \citet{Paardekooper10} and a 3-d prescription from \citet{JimenezMassset17}. The choice of torque prescription may alter or remove the location of migration traps depending on the disk model parameters chosen, or $M_{\rm SMBH}$), as outlined in \citet{Grishin23}. Embedded objects migration according to a migration torque calculated as 
\begin{equation}
 \Gamma_{\rm mig} = C \left(\frac{H}{r} \right) \Gamma_{0}    
\end{equation}
where 
\begin{equation}
    \Gamma_{0}= q^{2} \Sigma r^{4} \Omega^{2} \left( \frac{H}{r}\right)^{-3}
\end{equation}
where $q$ is the mass ratio $q=m_{\rm bh}/M_{\rm SMBH}$ of a BH of mass $m_{\rm BH}$, located in the disk at distance $r$ from the SMBH, with Keplerian orbital frequency $\Omega$, local disk surface density ($\Sigma$) and local disk aspect ratio ($H/r$) and $C$ is the coefficient of the migration torque prescription as chosen by the user.


Embedded black holes may also be expected to produce feedback on the surrounding gas and modify the migration torque. We provide a switch for users {\texttt{flag\_thermal\_feedback}} to test feedback modification to the migration torque; if feedback is on, the Paardekooper or Jim\'enez-Masset torque prescriptions are modified to include thermal feedback effects. We modify the Paardekooper torque prescription according to the formulation from \citet{Hankla20} as
\begin{equation}
    \frac{\Gamma_{\rm heat}}{\Gamma_{0}} \approx 0.05\frac{n \eta}{\gamma(\gamma -1)^{1/2}}\left(\frac{H}{r}\right)\frac{c \kappa^{1/2}}{\sigma^{3/2}}\frac{H^{2}\Sigma^{2}\Omega^{7/2}}{T^{6}}
\label{eq:heat}
\end{equation}
where $n$ is the index of $d\ln(P)/d(\ln r)$, $\gamma$ is the adiabatic index, $\kappa$ is the disk opacity, $\sigma$ is the Stefan-Bolzmann constant, $H$ is the disk height, $r$ the disk radius, $\Sigma$ the disk surface density, $\Omega$ the orbital frequency and $T$ the disk temperature. The Jim\'enez-Masset torque prescription is modified by thermal feedback as outlined in \citet{Grishin23}.

 We assume $\eta=1$ (Eddington accretion) and $\alpha=0.01$ as defaults. An interesting feature of this modification to the migration torque is that $\Gamma_{\rm heat}$ generally acts outwards, so in optically thinner regions of disk models, migration can be directed outwards as opposed to (typically) inwards. Torque saturation inhibits runaway outward migration in low-opacity disks (e.g. the outer regions of \citealt{TQM05}) using the \citet{Hankla20} approximation).

To ensure we will prevent overshooting of migrators in the presence of a possible migration trap (due to timesteps being longer than typical migration times) we allow the user to declare the existence of a migration trap at a set radius. If the migration trap radius is set to zero, no corrections to migration behavior are applied; however, if the migration trap radius is set to a value, and a migrator has tried to cross the trap radius in a given timestep, the semi-major axis of the orbiter is set to the value of the trap radius. Note this is not self-consistently calculated and must therefore be determined a priori by the user, ideally using a methodology that follows \cite{Bellovary16} or \cite{Grishin23}, whereby trap locations are calculated as a function of $M_{\rm SMBH}, \alpha, \dot{M}$. One of our future goals for the code is to set up this machinery so that traps and anti-traps are calculated automatically for the user.

In any case, independent of the reality of traps, a migration `swamp' or `traffic jam' should naturally be expected as disk surface density changes in the inner disk as we shall see below.

Embedded black holes are assumed to accrete at a rate the user can specify in terms of the Eddington rate. We use this accretion rate to determine the change in mass, spin magnitude and spin angle for each BH. Accretion leads to spin-up for prograde circularized orbiters. Gas accretion is assumed to have orbital angular momentum parallel to that of the disk ($L_{\rm disk}$), so circularized BH are assumed to torque towards alignment with $L_{\rm disk}$ over time. The default assumption is that prograde, circularized BH are torqued into alignment with $L_{\rm disk}$ once a user specified fraction of the BH mass has accreted (e.g. \citealt{Bogdanovic07} suggests $0.01$ to $0.1$ are reasonable values), and we assume a BH spins up from $a=0$ to $a=0.98$ once it accretes a factor $\sqrt{3/2}$ in mass \citep{Bardeen70}. We assume black holes on sufficiently eccentric orbits \citep{Chen22}
 \begin{equation}
     e > (h/r) \sqrt{(1+\lambda^{2})\,{\rm max}[1,3^{1/3}(q^{1/3}/(h/r)^{2})]-1}
 \end{equation}
 where $\lambda \sim 1.3$, $q=m_{\rm BH}/M_{\rm SMBH}$, will acrete retrograde from the gas and spin down, torquing away from $L_{\rm disk}$. 

We then compute the evolution of single, initially embedded retrograde sBH. The evolution of their semi-major axis, eccentricity, and inclination angle due to their interaction with the gas disk is handled in a single function which considers the complex interactions between these parameters. Retrograde orbiters evolve remarkably quickly into inner disk objects, if their argument of periapse, $\omega$ is near 0 or 180 degrees, while their orbits remain extremely stable over long periods if $\omega=90$ or $270$ degrees. To properly compute their trajectories, ideally we would use an N-body code such as \texttt{SpaceHub}, as demonstrated in \cite{WZL24}. However, such a strategy is computationally expensive, and so we instead use a crude approximation of their expected behavior, based on several key examples computed in \cite{WZL24}. We assume orbiters have $\omega$ either 0 or 90 degrees to cover the limiting cases, and for $\omega=0^{o}$ we assume their behavior follows that of the example shown in \cite{WZL24} Figure 12, where they compute the exact evolution for a black hole of $m_{\rm BH} = 30\,M_{\odot}$, $i_0=175^{o}$, $e_0=0.7$, $a_0=100\,r_g$ around a $M_{\rm SMBH}=10^8 \,M_{\odot}$ black hole, with an assumed surface density profile that matches \cite{SG03}.For such orbiters, they show 3 periods of evolution:

First, an orbiter will radialize (increase eccentricity) at the fastest timescale, decrease its semimajor axis (more slowly), and decrease its inclination angle (very slowly). Second, once the orbiter has reached a very large (nearly radial) eccentricity, it will decrease its inclination angle (very quickly), flipping towards a prograde orbit quite suddenly, while now decreasing its eccentricity and circularizing (more slowly than the prior radializing timescale), and doing this at approximately constant semimajor axis. Third, the inclination angle will continue to decrease extremely quickly until it is identically zero, while also continuing to circularize and shrink its semimajor axis (these latter two processes are slower than migration of circular prograde objects).

We retrieve the timescales and changes in parameters for each phase of this evolution from Figure 12 (specifically, stage 1 is $\sim1.5\times10^5$~yr, when $e=0.7\rightarrow 0.9999$, $a=100\rightarrow60~r_g$, $i=175\rightarrow165^{o}$; stage 2 is $\sim10^4$~yr, when $i=165\rightarrow12^{o}$, $e=0.9999\rightarrow0.9$, $a=60~r_g$; and stage 3 is $\sim10^4$~yr, when $i=12\rightarrow0.0^{o}$, $e=0.9\rightarrow0.5$, $a=60\rightarrow20~r_g$). Then, for any object in McFACTS, we scale the timescales for each phase according to Eqns 70-73 in \cite{WZL24}:

\begin{equation}
    \tau_{\rm p,dyn} = \frac {{\rm sin}~i~(\delta - {\rm cos}~i)^{3/2}}{\sqrt{2} \kappa ~| {\rm cos}~i - \zeta |} \frac{M_\mathrm{tot}}{m_{\rm BH}} \frac{M_\mathrm{tot}}{\Sigma \pi p^{2}} T
\end{equation}
\begin{equation}
    \tau_{\rm i,dyn} = \frac {\sqrt{2} i~(\delta - {\rm cos}~i)^{3/2}}{\kappa} \frac{M_\mathrm{tot}}{m_{\rm BH}} \frac{M_\mathrm{tot}}{\Sigma \pi p^{2}} T
\end{equation}
\begin{equation}
    \tau_{\rm a,dyn} = \frac {(1-e^2)~{\rm sin}~i~(\delta - {\rm cos}~i)^{3/2}}{\sqrt{2} \bar\kappa~| {\rm cos}~i - \bar\zeta |} \frac{M_\mathrm{tot}}{m_{\rm BH}} \frac{M_\mathrm{tot}}{\Sigma \pi p^{2}} T
\end{equation}
\begin{equation}
    \tau_{\rm e,dyn} = \frac{2e^2}{(1-e^2)} \left| \frac{1}{\tau_{\rm a,dyn}}-\frac{1}{\tau_{\rm p,dyn}} \right|^{-1}
\end{equation}
i.e. the timescales on which the parameters $p$, $i$, $a$, and $e$ change assuming only dynamical friction forces apply. Here $p=a(1-e^2)$ is the semi-latus rectum of the orbit, $T=2\pi \sqrt{a^3/GM_{\rm SMBH}}$ is the Keplerian period, $M_{tot}=M_{\rm SMBH} + m_{\rm BH}$, $\Sigma$ is the local mass surface density at $a$ and we also define:
\begin{equation}
    \sigma_{\pm} = \sqrt{1+e^2 \pm 2e~ {\rm cos} ~\omega}
\end{equation}
\begin{equation}
    \eta_{\pm} = \sqrt{1 \pm e~ {\rm cos}~ \omega}
\end{equation}
\begin{equation}
    \delta = \frac{1}{2} \left( \frac{\sigma_{+}}{\eta^2_{+}} + \frac{\sigma_{-}}{\eta^2_{-}} \right)
\end{equation}
\begin{equation}
    \kappa = \frac{1}{2} \left( \sqrt{\frac{1}{\eta_{+}^{15}}} + \sqrt{\frac{1}{\eta_{-}^{15}}} \right)
\end{equation}
\begin{equation}
    \xi = \frac{1}{2} \left( \sqrt{\frac{1}{\eta_{+}^{13}}} + \sqrt{\frac{1}{\eta_{-}^{13}}} \right)
\end{equation}
\begin{equation}
    \zeta = \frac{\xi}{\kappa}
\end{equation}
\begin{equation}
    \bar\kappa = \frac{1}{2} \left( \sqrt{\frac{1}{\eta_{+}^{7}}} + \sqrt{\frac{1}{\eta_{-}^{7}}} \right)
\end{equation}
\begin{equation}
    \bar\xi = \frac{1}{2} \left( \sqrt{\frac{\sigma_{+}^4}{\eta_{+}^{13}}} + \sqrt{\frac{\sigma_{-}^4}{\eta_{-}^{13}}} \right)
\end{equation}
\begin{equation}
    \bar\zeta = \frac{\bar\xi}{\bar\kappa}.
\end{equation}

For objects with $\omega=90$~degrees, the evolution is qualitatively different, with monotonic behavior in $a$, $e$, and $i$, such that the semimajor axis will decrease quite slowly, while the eccentricity will circularize even more slowly, and the inclination angle will decrease slowest of all. We scale the behavior for an orbiter identical to the first example, except for $\omega$, using scaling for timescales derived from \cite{WZL24} Figure 8 (specifically, in $\sim1.5 \times10^7~$yr $a=100\rightarrow60\,r_g$, $e=0.7\rightarrow0.5$, $i=175\rightarrow170^{o}$). We do not yet compute dynamical interactions between retrograde orbiters and prograde orbiters (or indeed between retrograde orbiters), as the timescales for their interactions are quite short---half of the initially embedded retrograde orbiters evolve into EMRIs in the first few timesteps (i.e. those with $\omega=0$~degrees), while the remainder evolve so slowly they are relatively unlikely to encounter any individual prograde object, which is typically moving more quickly through the disk. However, rare encounters could be extremely high energy, and we plan to add such interactions in future versions.

For prograde single orbiters which have been circularized by the gas, we now account for possible eccentricity re-excitation due to dynamical interactions. For any circular orbiters, we compute the odds of interacting with a co-planar eccentric orbiter. First we establish which of the eccentric BH population have periapse or apoapse that straddle a given circular orbiter. Then, we treat the average probability of encounter between a circularized BH of mass $M_{1}$ and eccentric interloper of mass $m_{2}$ simply as 
\begin{equation}
P_{\rm enc}=\frac{2R_\mathrm{H}}{\pi a}\frac{N_{\rm orb}}{\Delta t}
\end{equation}
where $R_{\rm H}$ is the mutual Hill sphere 
\begin{equation}
    R_{\rm H} = a\, (q/3)^{1/3}
\end{equation}
 with $a$ is the semi-major axis of $M_{1}$, $q=M_{1}+m_{2}/M_{\rm SMBH}$, and ($N_{\rm orb}/\Delta t$) is the number of orbits of $M_{1}$ per timestep($\Delta t$). We assume an average fractional energy change ($\Delta E \sim 0.1$, default) per strong encounter. We draw a probability from a uniform distribution $0<P<1$ and if $P<P_{\rm enc}$ then an encounter occurs. For each encounter we change circularized BH orbits from $(a_{1},e_{1})$ to $(a_{1}(1+\Delta E),e_{1}(1+\Delta E))$ and eccentric BH properties from $(a_{2},e_{2})$ to $(a_{2}(1-\Delta E),e_{2}(1-\Delta E))$. If $e_{1}(1+\Delta E)>e_{\rm crit}$ then we remove the excited BH from the circularized population and add it back to the eccentric population.

\subsubsection{Binaries}
Finally, we address the physics of binaries. On the first timestep in \texttt{v.0.3.0} there are zero binaries, and so all binaries are dynamically formed; however, for any future timestep we proceed as follows for any binaries that exist:

First, we damp $e_\mathrm{orb}$ due to gas interactions, using the same formulation as for single BH but assuming the mass is now $M_{\rm bin}=M_{1}+M_{2}$.

Then, we dynamically perturb the binary, first due to possible encounters with circular, co-planar single orbiters. We follow the same procedure for single perturbations described above, except we keep track of the relative energy ($E_{\rm rel}$) of an actual close encounter
 \begin{equation}
     E_{\rm rel} =\frac{\left(\frac{GM_{1}M_{2}}{a_{\rm b}}\right)}{\frac{1}{2} m_{3}v_{\rm rel}^{2}}
 \end{equation}
where $M_{1},M_{2}$ are the BBH component masses, $a_{b}$ is the semi-major axis of the BBH around its own center of mass, $m_{3}$ is the mass of the interloper and $v_{\rm rel}$ is the relative velocity of encounter. We calculate $P_{\rm enc}$ with $R_{H}$ the mutual Hill sphere of the BBH plus interloper and $a$ is now the BBH semi-major axis and $N_{\rm orb}$ the number of orbits of the BBH around the SMBH per timestep ($\Delta t$).

 For $E_{\rm rel}>1$ we assume a hardening encounter within the Hill sphere. For circularized BBH encountering circularized prograde BH, we allow the user to set the mean fractional energy exchange during an encounter as well as the  variance of the distribution of fractional energy exchange during the encounter. A full study of dynamical encounters (including typical numbers of hardening encounters) as a function of the stalling of gas-hardening at different binary separations will follow in Postiglione et al. 2025 (in prep.). 
 
 Post tertiary encounter, ($a_{b},e_{b}$) becomes ($a_{b}(1-\Delta E_{\rm strong}), e_{b}(1+\Delta E_{\rm strong})$)  and $e_{\rm orb,b}$ becomes $e_{\rm orb,b}(1+\Delta E)$. The circularized BH interloper changes from ($a_{3},e_{3}$) to ($a_{3}(1+\Delta E_{\rm strong}),e_{3}(1+\Delta E_{\rm strong}))$. Thus, in this setup, a sufficiently hard BBH population can rapidly be driven to merger via multiple (but few) dynamical encounters at low relative velocity with circularized prograde BH, following \citet{Leigh18}. The default energy exchange for circularized encounters is drawn from a distribution centered on $\Delta E_{\rm exch}= 0.9 \pm 0.025$. A drop to $E_{\rm exch} =0.5$ or even $E_{\rm exch}=0.1$ does not substantially alter the rate of mergers, dropping the rate by $\sim 1-3\%$. This is because for our default model, gas hardening dominates and runs away at small binary separation as we have yet to implement gas stalling. Once gas stalling is implemented, the energy exchange distribution will become far more important for BBH hardening and mergers. For example, we note that mergers near the migration trap do appear to be inhibited more for lower values of $\Delta E_{\rm exch}$. (see Postiglione et al. 2025 (in prep.). For $E_{\rm rel} <1$, a softening encounter happens and ($a_{b},e_{b}$) becomes ($a_{b} (1 + \Delta E),e_{b}(1-\Delta E)$ and $e_{\rm orb,b}$ becomes $e_{\rm orb,b}(1+\Delta E)$. The circularized interloper is assumed to change from ($a_{3},e_{3}$) to ($a_{3}(1+\Delta E),e_{3}(1+\Delta E)$). 
 
 Then we allow interactions between circularized binaries and eccentric co-planar single orbiters following the same process as for single circularized BH above, except $\Delta E_{\rm strong}$ above becomes $\Delta E$ (i.e we only allow rapid dynamic hardening for mutually circularized encounters at low relative velocity). We ignore encounters between eccentric BBH and eccentric BH for now and only circularized BBH can migrate within the disk. 

Next, existing binaries can be hardened by gas interactions and we assume that they obey the gas hardening prescription found by \cite{Baruteau11}, unless their expected merger time due to GW emission \citep[computed using][]{Peters64} is shorter. 

Then, we update the binary component parameters of mass, spin, and spin angle, due to accretion, and we use the same prescriptions as for single objects.

In addition, we allow for dynamical interactions between the embedded BBH population and the spheroid population. We assume the default spheroid encounter is with a star of mass $2M_{\odot}$ and we assume that within 1Myr, all the orbits of stars within $a<10^{3}r_{g}$ are removed from the spheroid and captured by the disk (although embedded stars are not included in \texttt{v.0.3.0}). $P_{\rm enc}$ is calculated in a manner similar to above, as a function of the location of the BBH in the disk ($a$) and the density of the disk and spheroid populations respectively. 

To allow for mergers that occur rapidly out of plane, such that $\chi_{p}>0$, before potentially rapid recapture, we assume $\Delta E_{\rm strong}=0.9$ for hardening encounters as for the in-disk circularized encounters, although this can be changed by the user. For spheroid encounters, we calculate $P_{\rm enc}$ using the density of spheroid encounters from \citet{Leigh18} (see their Fig.~1 for the approximate rate of spheroid encounter as a function BBH location ($a$) and BBH size ($a_{b}$)). For close spheroid encounters, we draw an angle of encounter from a uniform distribution of inclination angles. Since the spheroid population is captured by the disk over time \citep[e.g.][]{Rowan25}, we assume that low inclination angle orbiters are captured by the disk first \citep{Whitehead25b}. Thus the probability of very small inclination encounters vanishes early on (depending on disk model density) and the probability of slightly larger inclination encounters drops as $\tau_{\rm AGN}$ persists. The code chooses spheroid encounter angles at disk radius $R$ from a uniform distribution beginning at $i=h/R$. At present we assume that the lower bound to this distribution (i.e. the angle of encounter closest to the disk increases slowly each timestep at a rate that corresponds to the capture rate of objects in \citep{Fabj20}, see also below.

Since interactions with the spheroid population allow binaries to gain non-zero values of $i_{\rm orb,b}$ we allow gas damping of such binaries to return them to the mid-plane. In \texttt{v.0.3.0.} we estimate the time to disk recapture simply following \citet{Fabj20} where for a default \citet{SG03} disk, if $i_{\rm orb,b}<5^{\circ}$, the time to midplane recapture is $\sim 0.25 \,{\rm Myr} (M_{b}/40\,M_{\odot})^{-1}(a_{b}/10^{4}\,r_{g})$ and if $5^{\circ}<i_{\rm orb,b}<15^{\circ}$ , the time to midplane recapture is $\sim 12{\rm Myr}(M_{b}/40\,M_{\odot})^{-1}(a_{b}/10^{4}\,r_{g})$. We allow such binaries to continue to gas-harden per timestep, as they continue to accrete as they travel through the disk for typically $>1/2$ an orbit. Thus, they can merge quickly out of plane before midplane recapture. A more sophisticated treatment will be applied in future work. 

Finally, we migrate binaries (with sufficiently small $i_{\rm orb,b}$ and $e_{\rm orb,b }$) using the same Type 1 migration prescription described for single orbiters, but using the total binary mass in place of the single orbiter mass and binary center of mass to the semi-major axis, then applying our feedback model (above) to the net torque.

At this point, due to various processes, we check for any BBH mergers, and compute the strain and frequency of all mergers as well as existing BBH (here we assume a uniform distance $z=0.1$ by default for plots (see below), but users can rescale as appropriate). We also check for ionization of binaries. For mergers, we compute the output merger parameters including $\chi_{\rm eff}$ and remnant mass \citep{Tichy08} and reassign the newly formed remnant to the single BH category. We compute a kick at merger for a binary of mass ratio ($q=M_{1}/M_{2} \leq 1$), asymmetric mass ratio ($\nu = q/(1+q)^{2}$) and associated spin magnitudes ($a_{1},a_{2}$) following  \citep{Akiba24} as 
\begin{equation}
{\overrightarrow{v}}_{\rm kick} = v_{m}\hat{x} + v_{\perp}(cos\xi\hat{x} + sin\xi\hat{y}) + v_{\parallel}\hat{z}
\label{eq:v_kick}
\end{equation}
where $\hat{x},\hat{y},\hat{z}$ are the orthogonal unit basis vectors.  $v_{m}$ is a velocity given by
\begin{equation}
    v_{m}= A\nu^{2}\sqrt{1-4\nu}(1+B\nu)
\end{equation}
where the perpendicular ($v_{\perp}$) and parallel ($v_{\parallel}$)  components of the kick velocity (relative to the binary orbital angular momentum) are given by
\begin{equation}
    v_{\perp}=\frac{H\nu^{2}}{1+q}(a_{1,\parallel} - 
    q a_{2,\parallel})
\end{equation}
and
\begin{eqnarray}
    v_{\parallel}&=&\frac{16\nu^{2}}{1+q}[V_{1,1} + V_{A}\overrightarrow{S}_{\parallel} + V_{B}\overrightarrow{S}^{2}_{\parallel}+ V_{C}\overrightarrow{S}^{3}_{\parallel}] \nonumber\\
    &\times&|a_{1,\perp} -q a_{2,\perp}|cos(\Delta\phi)
\end{eqnarray}
where $\Delta\phi$ is a phase angle assumed to be uniformly distributed between $[0,2\pi]$ and where $H,V_{1,1},V_{A}, V_{B}, V_{C}$ are all constants from \citet{Akiba24} and with $\overrightarrow{S}$ defined as
\begin{equation}
    {\overrightarrow{S}}=2 \frac{\overrightarrow{a}_{1}+q^{2}{\overrightarrow{a}}_{2}}{(1+q)^{2}}.
\end{equation}
A kick can put a newly merged BH back in the eccentric BH population and we remove the binary; similarly we reassign ionized binary components and remove the binary, as appropriate.

Finally, we allow for the creation of new binaries. We assume that BH binaries (BBH) can form once they encounter each other on circularized prograde orbits ($e<e_{\rm min,crit}$) within $R_{\rm H}=a_{\rm bin,orb}(q/3)^{1/3}$ with $q=M_{\rm bin}/M_{\rm SMBH}$ and $a_{\rm orb}$ is the mass-weighted binary center of mass. We intend to incorporate a wider range of binary formation conditions \citep{Stan23,Rowan23,Qian24,Whitehead24} in future versions of the code. Note that all binaries are required to be on initially circular orbits around the SMBH, since we do not permit binary formation for non-circular objects.

We note that the removal of objects from the inner spheroid (as the spheroid is depleted due to disk interactions) corresponds to disk captures of new objects---we assume BH are captured by this disk model at a fixed rate of $1/0.1$Myr \citep{Secunda21}, drawn uniformly from the IMF distribution and deposited at small disk radii ($<2000\,r_{g}$)
 \citep{Fabj20}. They are added after the creation of new BBH and before inner disk processes.

\subsubsection{The inner disk}
Given both migration and disk capture have now occurred, we now check for `inner disk objects' defined by the user (default $<50 \,r_{g}$). For such objects in \texttt{v.0.3.0.} we assume pure GR evolution \cite{Peters64}. We compute the GW strain and frequency for such objects in orbit around the SMBH (subject to the same assumptions as for the BBH above). Then we check for any actual mergers with the SMBH and update our filing cabinet to remove merged objects.

At this point we also check for objects that may have flipped fully from retrograde to prograde (due to the single evolution discussed above), reclassify them as necessary and update the filing cabinet.

Finally, we reach the end of the timestep and repeat until the number of timesteps (or equivalently $\tau_{\rm AGN}$, requested by the user has been reached.



\subsection{Outputs}
For every iteration of a run (termed a galaxy) up to \texttt{galaxy$\_$num} we write \verb|initial_params_bh.dat| and \verb|output_mergers.dat|. After all runs we write \verb|out.log|, containing all input assumptions as well as the four key data files: \newline
1) \verb|output_mergers_emris.dat|, \newline 2) \verb|output_mergers_lvk.dat|, \newline 3) \verb|output_mergers_population.dat|, and \newline 4) \verb|output_mergers_survivors.dat|, corresponding to lists of properties of 1) EMRIs, 2) BBH with GW frequencies in the LISA/LVK bands, 3) BBH mergers, 4) All BH (singles and binary components) at the end of the AGN, all as a function of \texttt{galaxy} and \texttt{time}.

Based on these four key data files we presently generate the following standard plots using \texttt{make plots} (with examples of the default output plots in brackets after each): 1) Number of BH mergers as a function of remnant mass and BH generation (e.g. Fig.~\ref{fig:mf2}), 2) $M_{1}$ vs $M_{2}$ as a function of BH generation (e.g. Fig.~\ref{fig:m1m2}), 3) Mass of BH merger mass as a function of disk radius and generation (Fig.~\ref{fig:dyn_ecc}, left panel), 4) Mass of BH merger mass against time of merger and BH generation (Fig.~\ref{fig:dyn_ecc}, right panel), 5) $\chi_\mathrm{p}$ as a function of disk radius and generation (Fig.~\ref{fig:chi_p}, left panel), 6) BH mass ratio ($q=M_{2}/M_{1}$) as a function of $\chi_{\rm eff}$ and BH generation (Fig.~\ref{fig:qX}, left panel) , 7) Kick velocity distribution as a function of disk radius and BH generation (Fig.~\ref{fig:v_kick}) and 8) GW frequency characteristic strain per frequency per year vs average GW frequency (Fig.~\ref{fig:gw_strain}, left panel).

\section{Default Model Assumptions}
\label{sec:default}
In the results that follow, default model parameters (also in \texttt{p1$\_$model$\_$default.ini}) are listed in Table~\ref{tab:default_model}. Unless otherwise specified, we assume a \citet{SG03} disk model as interpolated from the \texttt{pAGN} package \citep{pAGN24} around a $10^{8}\,M_{\odot}$ SMBH, spanning $[6,50000]\,r_{g}$, with disk viscosity parameter $\alpha=0.01$ and lasting $0.5$ Myr. Average disk aspect ratio for this model is assumed to be $\sim 0.03$ and EMRIs are tracked within $50\,r_{g}$ of the SMBH. A migration trap is assumed to live at $700\,r_{g}$ (\citealt{Bellovary16}, although see also \citealt{Grishin23}). 

Following \citet{Generozov18}, a nuclear star cluster (NSC) is assumed to fill the inner galactic nucleus out to $5$pc, with an inner number density profile $n\propto r^{-7/4}$ within $r\sim 0.25$pc and $n \propto r^{-2.5}$ outside that. The BH population of the NSC is normalized at $10^{4}$ inside the central ${\rm pc}^{3}$, matching inferences of the population in our own Galactic nucleus. The BH population mass distribution has a Pareto form with index $M^{-2}$, $M_{\rm BH, min}=10\,M_{\odot}$, $M_{\rm BH,max}=40\,M_{\odot}$. BH drawn from this distribution above $M_{\rm BH,max}$ are randomly assigned masses in a Gaussian pile-up centered on $35\,M_{\odot}$ of width $2.3\,M_{\odot}$, mimicking the observed pile-up in the LVK mass spectrum \citep{massbump23}. The initial BH spin distribution is assumed to be drawn from a Gaussian centered on dimensionless spin parameter $=0$ and of width $0.1$. Assuming an average scale height across the \citet{SG03} disk model, the fraction of BH with orbits coincident with the disk volume is calculated and the corresponding radial BH locations are generated from a distribution reflecting the user choice of NSC density.

Our default accretion rate assumptions for determining the change in mass, spin magnitude, and spin angle of each embedded black hole is that they accrete at Eddington, and must accrete $0.01$ of their own mass (while on a circular, prograde orbit) in order to have their spin aligned with the disk orbital angular momentum.
 
Our default assumption is a uniform distribution of eccentricity, i.e. sub-thermal, allowing for e.g. only partial recovery from a previous AGN episode that might have dynamically cooled these orbits. The uniform distribution is capped at some initial eccentricity value $e_{\rm max}$ to test the likely rate of relaxation from a previous AGN episode. A full range of initial eccentricity distributions will be tested in future work.

 We assume that gas hardening of BBH follows the prescription in \citet{Baruteau11}. Our default model allows for dynamical interaction between the circularized population of BH in the AGN disk and the eccentric population (i.e. dynamics flag is on). 

Finally, we estimate the approximate merger rate corresponding to a particular simulation, by assuming a number density of AGN $n_{\rm AGN} \sim 10^{-3} \,{\rm Mpc^{-3}} \sim 10^{6} \,{\rm Gpc^{-3}}$. This yields the very useful approximation~\footnote{ Dong Lai, private communication} that $\mathcal{R}=1$ merger/AGN/Myr in these simulations is equivalent to a rate of $\mathcal{R}=1\,{\rm Gpc^{-3} yr^{-1}}$. We caution this is only an approximate rate estimate, as we are not accounting for the variation in merger rates or the distribution of merger parameters that may occur with variations in SMBH mass, NSC mass and AGN disk model. Such complications are explored in companion paper \citep{Delfavero24}, where we attempt to simulate the AGN channel for the universe, with a distribution of galaxy properties.

\section{Testing models with {\texttt{McFACTS}}}

\subsection{Testing nuclear star cluster models}
\begin{figure*}
    \centering
    \includegraphics[width=1\columnwidth]{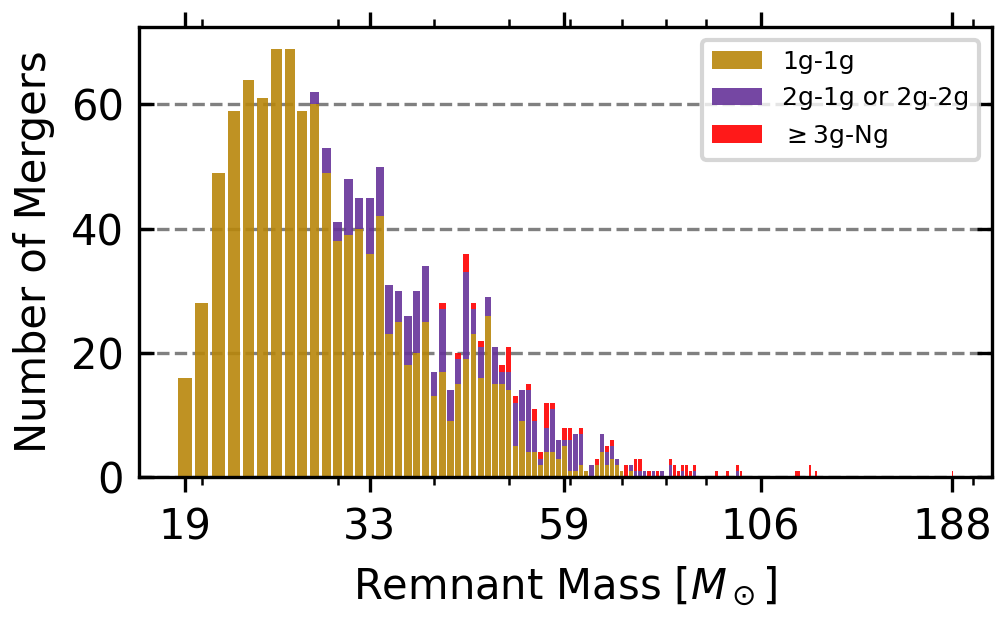}
      \includegraphics[width=1\columnwidth]{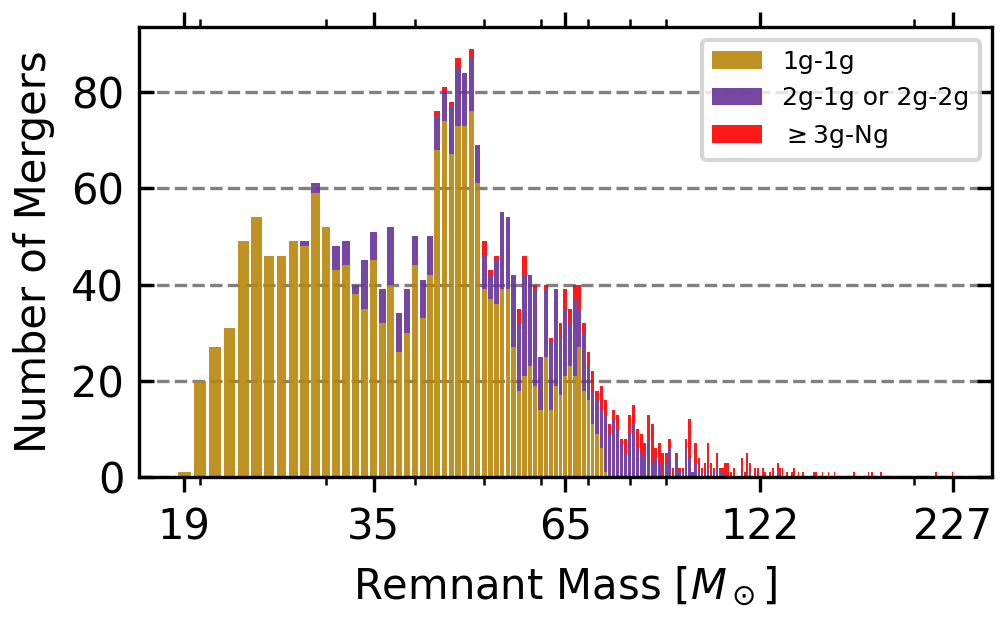}
    \caption{Number of BBH mergers per generation of BH as a function of merged mass ($M_\odot$) for different BH initial mass functions (IMFs). Left panel corresponds to a BH IMF of $M^{-2}$ with minimum $M_{\rm BH}=10\,M_{\odot}$, maximum initial mass $M_{\rm BH, max}=40\,M_{\odot}$ and simple pile-up model between $[35,40]\,M_{\odot}$ for a 0.5Myr, SG disk model around a $M_{\rm SMBH}=10^{8}\,M_{\odot}$ (see text). Gold denotes 1g-1g mergers, purple denotes 2g-mg ($m\leq 2$)and red denotes 3g-ng ($n \leq 3$) and higher, where 1g is first generation BH (not previously involved in a merger), 2g is second generation (the result of 1 merger) etc. Right panel is as left panel but with BH IMF of $M^{-1}$. Input file is \texttt{model$\_$choice$\_$old.ini} with \texttt{nsc$\_$imf$\_$bh$\_$powerlaw$\_$index = 2.(1.)} left (right), \texttt{galaxy$\_$num = 100} and \texttt{seed = 3456789018}. Equivalent merger rates correspond to $\mathcal{R}\sim 22.0(40.0) \,{\rm Gpc}^{-3} \rm {yr}^{-1}$ in left (right) panel.
   }
    \label{fig:mf2}
\end{figure*}

Using the default values outlined above, Fig.~\ref{fig:mf2} (left panel) shows the number of BBH mergers as a function of remnant mass.
In Fig.~\ref{fig:mf2} (left panel), the mass function of merged BH peaks between $[20,30]\,M_{\odot}$, with a secondary peak at $\sim [40,50]\,M_{\odot}$. The dominant, lower-mass peak is driven by mergers among the most numerous low mass BH in the IMF ($\sim [10,15]M_{\odot}+[10,15]\,M_{\odot}$). The second peak is driven by mergers between masses in the pile-up and the low masses ($\sim [35]\,M_{\odot} + [10-15]\,M_{\odot}$), since the former are the fastest migrators and the latter are the most common BH. The equivalent mean merger rate is $\mathcal{R}\sim 22.0\,{\rm Gpc}^{-3}{\rm yr}^{-1}$.

\begin{figure*}
    \centering
    \includegraphics[width=1\columnwidth]{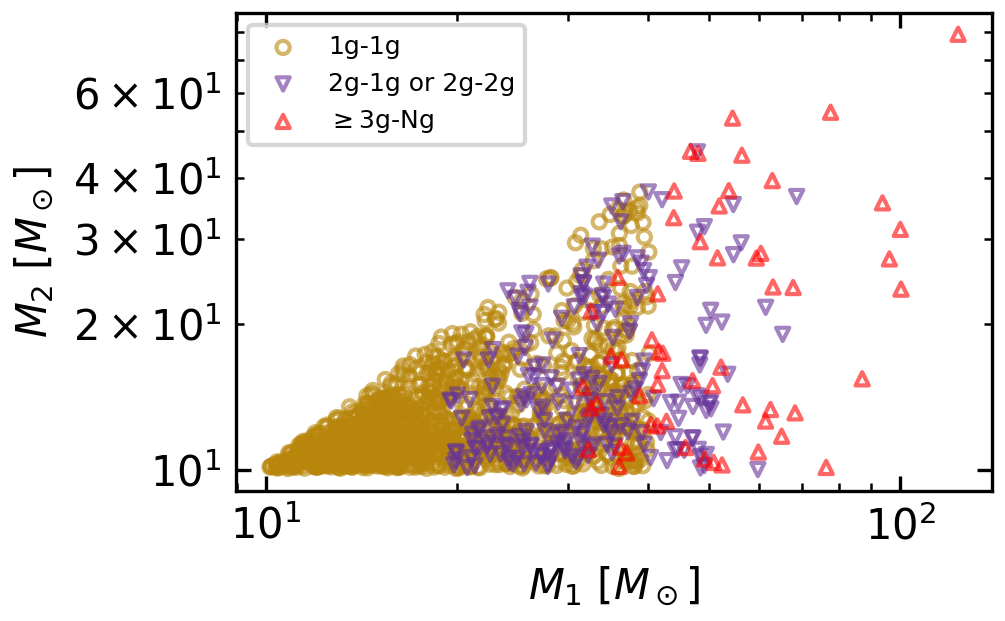}
      \includegraphics[width=1\columnwidth]{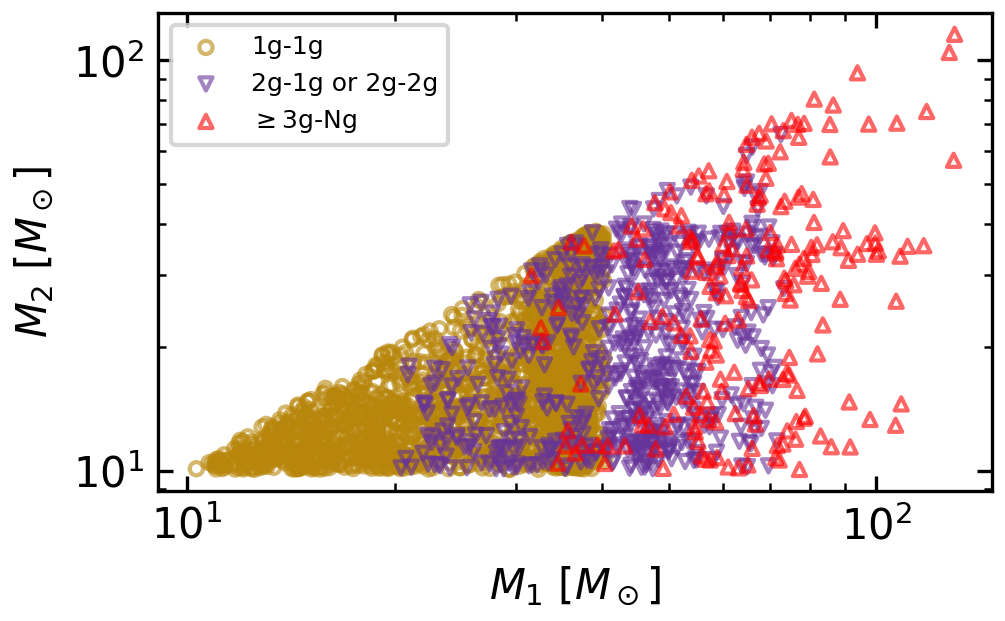}
    \caption{Distribution of BH masses ($M_{1},M_{2}$) involved in the mergers in Fig.~\ref{fig:mf2}.
   }
    \label{fig:m1m2}
\end{figure*}

By contrast, Fig.~\ref{fig:mf2} (right panel) shows the same quantities, for the same BH minimum and maximum mass and toy-model pile-up but with $M_{\rm BH} \propto M^{-1}$. Qualitatively we can see that a flatter IMF drives more numerous and more massive mergers typically, with a longer tail to high mass and a significant number of IMBH ($>100\,M_{\odot}$) produced. The equivalent mean merger rate is $\mathcal{R}\sim 40.0\,{\rm Gpc}^{-3}{\rm yr}^{-1}$. Here, a flatter IMF reverses the magnitudes of the two peaks in Fig.~\ref{fig:mf2} (left panel), with the peak between $\sim [45,55]\,M_{\odot}$ dominating the merger spectrum. We also observe the emergence of a third peak at $\sim [65,75]\,M_{\odot}$, followed by a long hierarchical tail to higher masses. The toy model pile-up between $[35,40]\,M_{\odot}$ is relatively more prominent for a $M^{-1}$ IMF so both this feature and mergers among BH in this feature (at $[70,75]\,M_{\odot}$)are significantly more prominent than in the left panel. The new, third peak is an echo of the pile-up in the IMF yielding BH around $\sim 70\,M_{\odot}$ due to $\sim 35\,M_{\odot}+35\,M_{\odot}$ mergers since the relatively numerous massive BH ($\sim 35\,M_{\odot}$) migrate fastest and tend to find each other more quickly. This picture is confirmed in Fig.~\ref{fig:m1m2} where we see an excess merger rate among the 1g-1g (gold) population in the right panel at $30-40$ in $M_{1}$ and indeed at ($M_{1},M_{2}$)$\sim 30-40\,M_{\odot}$ compared to the left panel. Intriguingly, there may be hints of an echo of the $35\,M_{\odot}$ peak in O3 LVK data at $\sim 70\,M_{\odot}$ which, if true, are suggestive of hierarchical mass mergers in a deep potential well \citep{Ignacio24}.  Observations of a relatively faint echo feature around $\sim 70\,M_{\odot}$ implies an IMF $\propto M^{[-1.5,-2]}$ for the AGN channel, but also requires a lower fraction of low mass BH to minimize intermediate features. Possibly a broken powerlaw IMF may be required as an input to AGN channel models. Or kicks may be important in segregation of generations between short-lived AGN phases. The relative positions and strengths of such peaks may be key in disentangling different BBH formation pathways in LVK data. For example, if in O4-O5, higher mass peaks are observed among $70M_{\odot}+70_{\odot}$ and $140_{\odot}+70_{\odot}$ BBH, these would be extremely suggestive of mergers retained in a very deep potential well, since such high generation and high spin mergers are very unlikely to be retained in clusters \citep[see also][]{Ford25}. The magnitude and relative sizes of such echo features in the mass spectrum will be explored more thoroughly in future work. 

\subsection{The effect of orbital eccentricity damping}
AGN disks damp the orbital eccentricity of prograde orbiters over time \citep[e.g.][]{McK12} but if the surface density increases with radius (as is the case in the inner region of the \citet{SG03} disk model) orbital eccentricities of retrograde orbits can be pumped \citep{Secunda21}.

Since $t_{\rm damp} \propto h^{4}$, orbital eccentricity damping is fastest in thin parts of disk models ( e.g. $\sim [10^{2},10^{4}]\,r_{g}$ in \citealt{SG03}). Therefore we expect BH orbits to circularize fastest in the thinnest parts of the disk. Since BBH formation is predominantly drawn from the population that have already circularized, we expect the BBH population to tend to form in the thinner parts of the disk model first.
 
If we assume an initially thermal distribution of orbital eccentricities, we expect an orbital distribution function 
\begin{equation}
    f(e){de} = 2e \;{de}
\end{equation}
such that the median orbital eccentricity is $e \sim 1/\sqrt{2} \sim 0.7$ and there is a uniform probability distribution of $e^{2}$. For initially thermal eccentricity distributions, we draw from a uniform distribution [0,1] to obtain $e^{2}$ at $t=0$. However, such a distribution is the limiting case where orbits have had sufficient time to relax and will only apply where there is a very long interval between AGN episodes. Instead, for our default case, which involves a short (0.5Myr) AGN lifetime, we introduce a maximum initial eccentricity ($e_{\rm max}$), capping the uniform distribution, corresponding to a relatively short period of relaxation between AGN episodes.

We expect new BBH to form mostly from the circularized population. The subsequent migration of such BBH either inwards or outwards, away from this disk region, will make dynamical encounters with eccentric orbiters more likely. Such encounters will be capable of either hardening, softening/ionizing the BBH, depending on the details of the encounter \citep[e.g.][]{Leigh18,Yihan21}. Note also $a \sim 10^{3}\,r_{g}$ is a plausible location for a migration trap \citep{Bellovary16} in such a disk, although \citet{Grishin23} suggest that such traps can occur at radii $\times 3-5$ further out in the disk.

\begin{figure*}
    \centering
    \includegraphics[width=1\columnwidth]{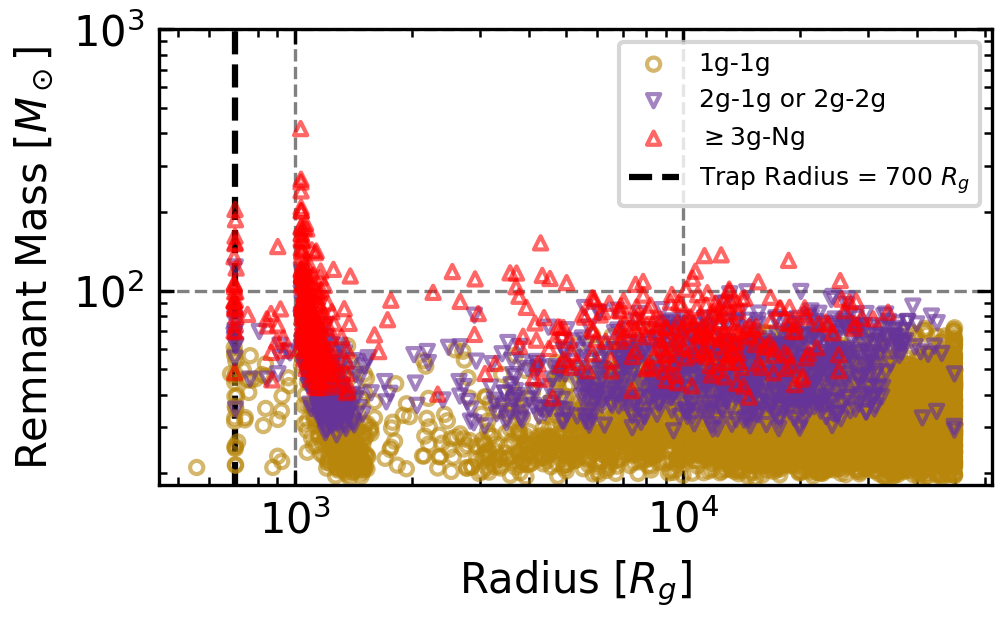}
      \includegraphics[width=1\columnwidth]{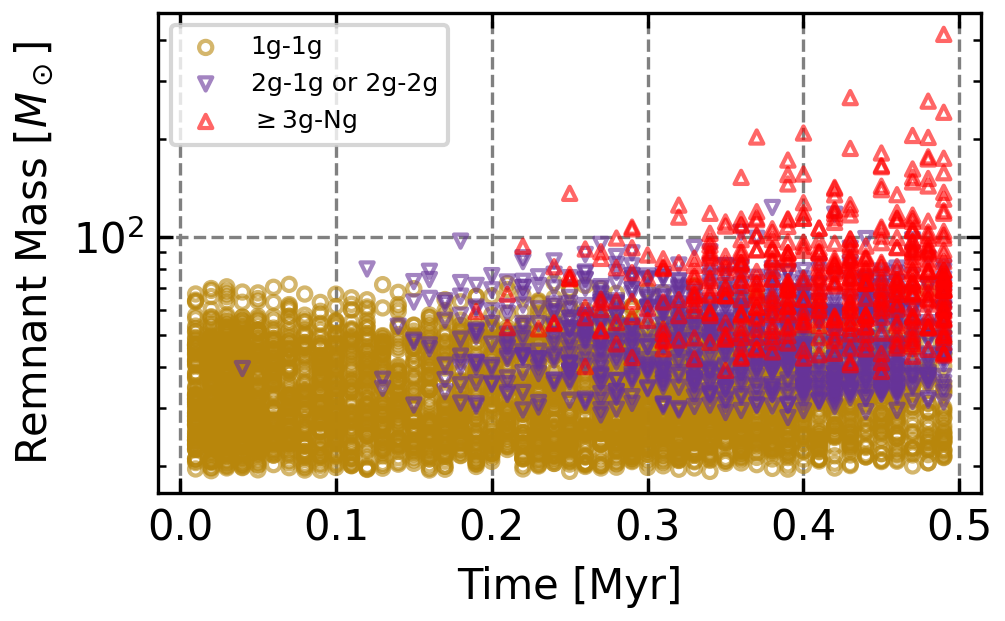}
    \caption{BBH merger mass as a function of disk radius (left panel) and time (right panel). All BH are assumed initially circularized and no dynamical encounters occur. Input file is \texttt{model$\_$choice$\_$old.ini} with \texttt{flag$\_$orb$\_$ecc$\_$damping = 0}, \texttt{flag$\_$dynamic$\_$enc = 0}, \texttt{nsc$\_$spheroid$\_$normalization = 0.0}, \texttt{galaxy$\_$num = 100} and \texttt{seed = 3456789018}. Equivalent merger rate is ${\mathcal{R}} \sim 95.4\, {\rm Gpc}^{-3} {\rm yr}^{-1}.$
   }
    \label{fig:circ}
\end{figure*}

Fig.~\ref{fig:circ} shows the BBH merger masses as a function of disk radius (left panel) and time (right panel) for the setup as in Fig.~\ref{fig:mf2}, except all the BH are assumed to be initially circularized and we do not allow dynamical interactions. Feedback is on, so migration develops an outward component in the outer disk (driving a small pile-up near the outer disk edge). The equivalent merger rate is very high $\mathcal{R} \sim 95.4\,{\rm Gpc}^{-3} {\rm yr}^{-1}$ as BBH mergers start straightaway ($t=0$, right panel) and high generation mergers (2g,3g) quickly appear in $<0.1$ Myr. There is a divergence in the mass distribution of mergers in the right panel with IMBH-building mergers ($>100M_{\odot}$) starting in the first $\sim 0.2$ Myr, and growing larger over time. These  IMBH mergers occur preferentially at both the migration trap \citep{Bellovary16}, and the migration swamp in front of the trap around $\sim 10^{3}r_{g}$, where the Paardekooper migration torque drops by $\sim 30\%$ for $M_{\rm SMBH} \sim 10^{8}M_{\odot}$, leading to a pile-up at this region (see e.g. Fig.2 in \citep{Grishin23} for an approximate illustration). The bulk of mergers are at lower mass, which was also the pattern we found in \citet{McK20a} where we assumed all the BH are on circularized orbits at $t=0$. 

The largest mass IMBH occur at or near the migration trap or at the disk outer boundary. Some IMBH-producing mergers (up to $\sim 150\,M_{\odot}$) occur far from the trap, in regions of the disk where the size of the BBH Hill sphere radius is large. 

\begin{figure*}
    \centering
    \includegraphics[width=1\columnwidth]{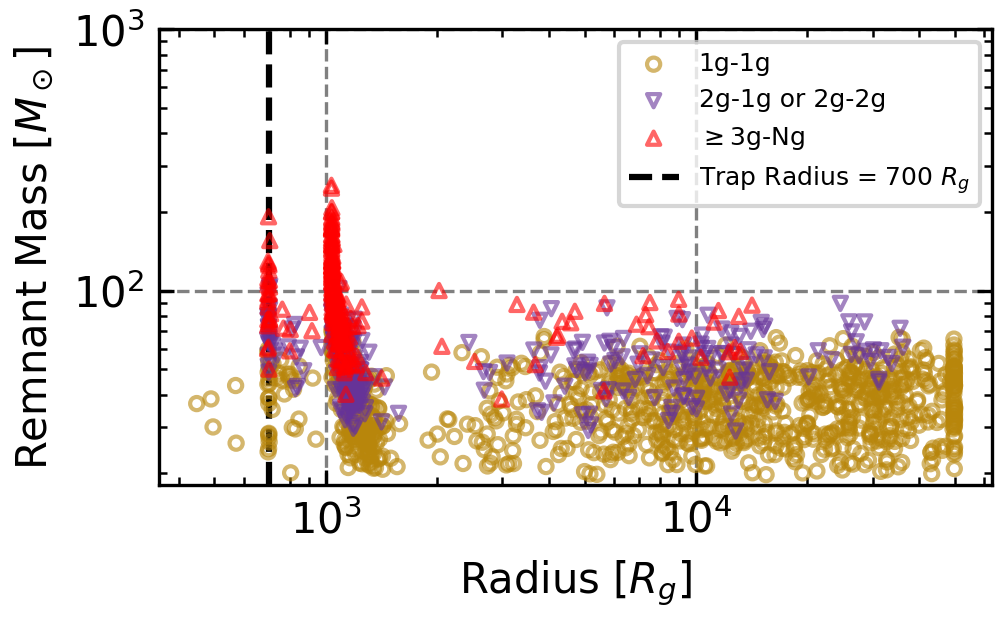}
      \includegraphics[width=1\columnwidth]{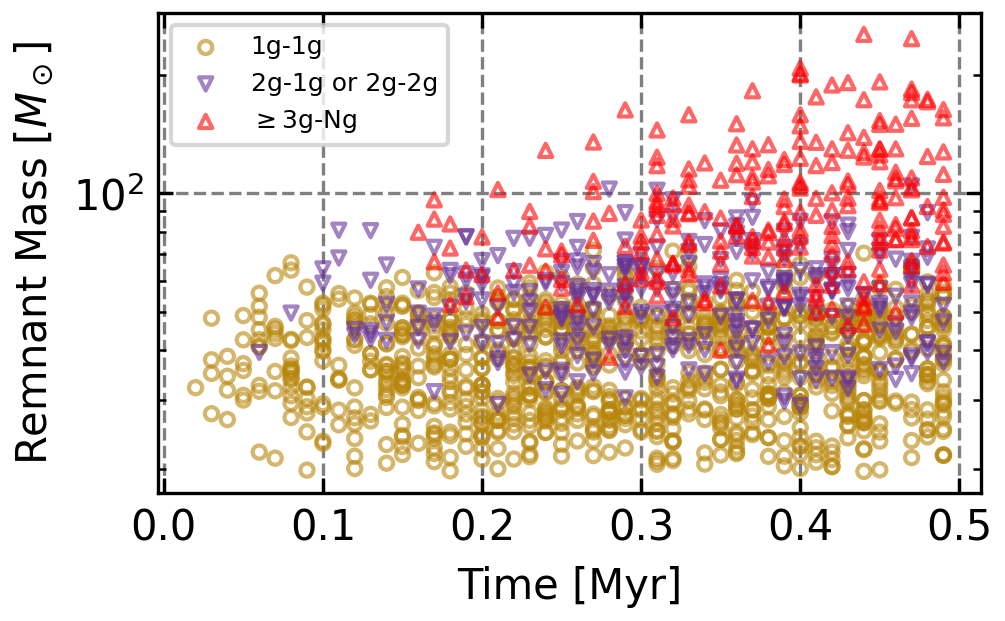}
    \caption{As Fig.~\ref{fig:circ} except all BH are assumed to be drawn from an initial uniform distribution of orbital eccentricities between [$0.0,0.9$] and no dynamical encounters occur. Input file is as Fig.~\ref{fig:circ}, except \texttt{flag$\_$orb$\_$ecc$\_$damping = 1} and \texttt{disk$\_$bh$\_$orb$\_$ecc$\_$max$\_$init = 0.9}. Equivalent merger rate is ${\mathcal{R}} \sim 26.8 \,{\rm Gpc}^{-3} {\rm yr}^{-1}.$
   }
    \label{fig:ecc}
\end{figure*}

\begin{figure*}
    \centering
    \includegraphics[width=1\columnwidth]{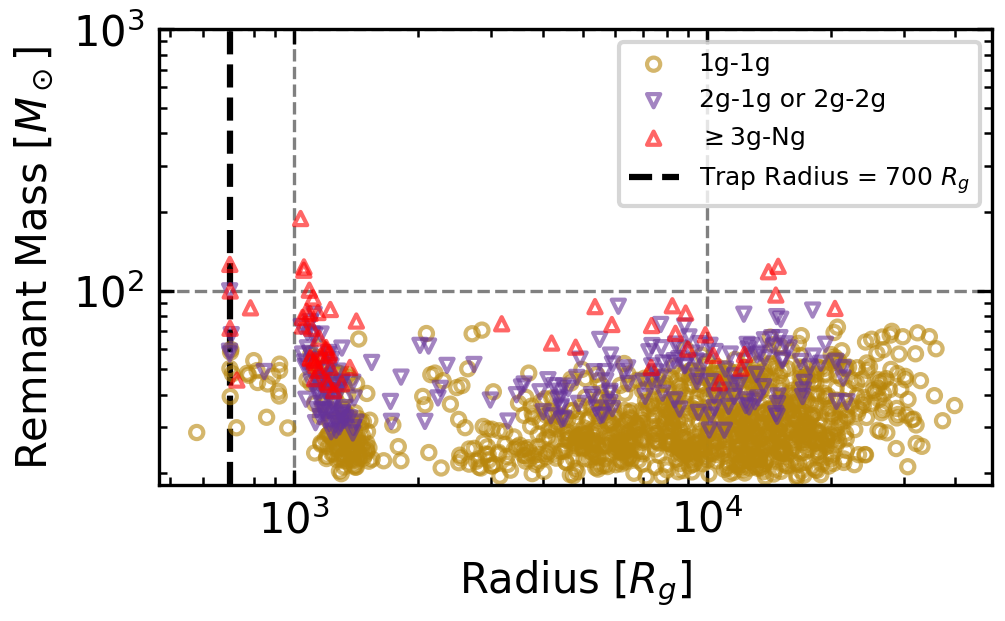}
      \includegraphics[width=1\columnwidth]{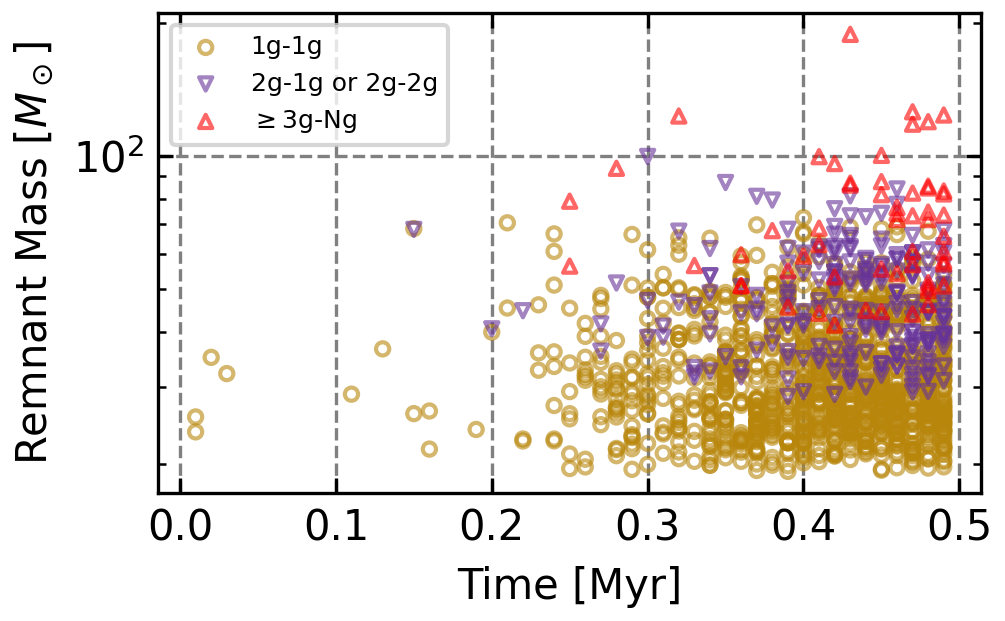}
    \caption{As Fig.~\ref{fig:ecc} except all BH are assumed to be drawn from an initial uniform distribution of orbital eccentricities between [$0.0,0.3$] and dynamical encounters are allowed. Input file is \texttt{model$\_$choice$\_$old.ini} with \texttt{flag$\_$orb$\_$ecc$\_$damping = 1}, \texttt{flag$\_$dynamic$\_$enc = 1}, \texttt{nsc$\_$spheroid$\_$normalization = 1.0}, \texttt{disk$\_$bh$\_$orb$\_$ecc$\_$max$\_$init = 0.3}, \texttt{galaxy$\_$num = 100} and \texttt{seed = 3456789018}. Equivalent merger rate is ${\mathcal{R}} \sim 22.0 \,{\rm Gpc}^{-3} {\rm yr}^{-1}.$
   }
    \label{fig:dyn_ecc}
\end{figure*}

\begin{figure*}
    \centering
    \includegraphics[width=1\columnwidth]{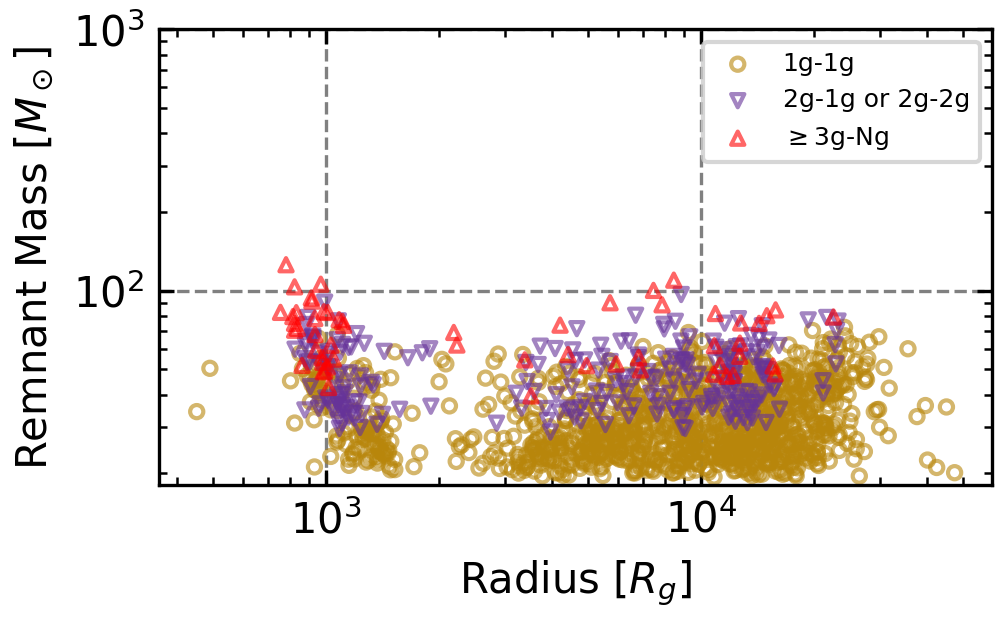}
      \includegraphics[width=1\columnwidth]{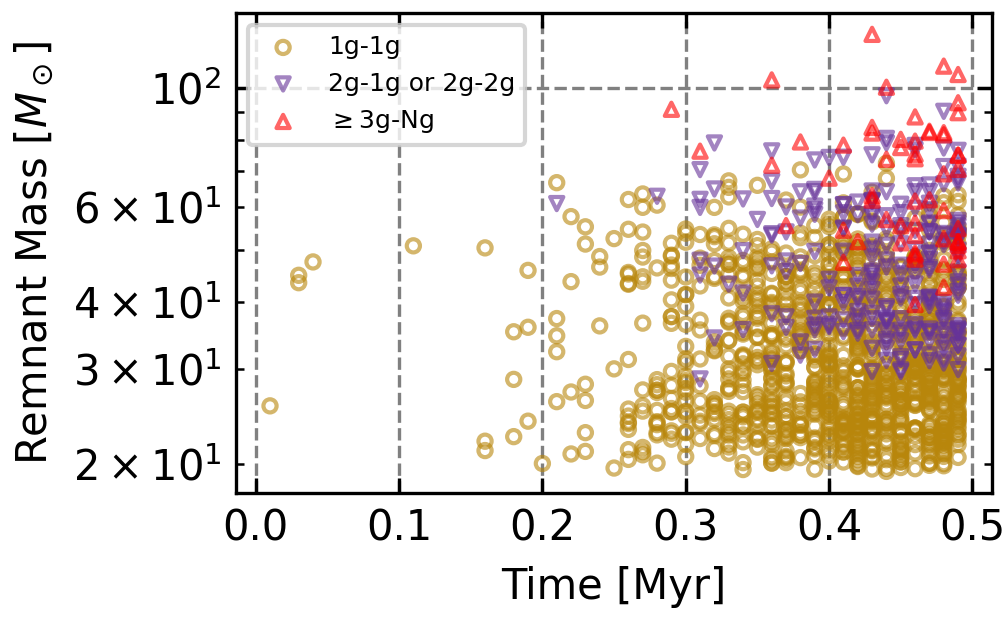}
    \caption{As Fig.~\ref{fig:dyn_ecc}, except  \texttt{torque$\_$prescription = jimenez$\_$masset}. Equivalent merger rate is ${\mathcal{R}} \sim 21.2\ {\rm Gpc}^{-3} {\rm yr}^{-1}.$
   }
    \label{fig:jm}
\end{figure*}

By contrast, Fig.~\ref{fig:ecc} shows what happens if we assume an initially eccentric population (\texttt{flag$\_$eccentricity$\_$damping = 1} with a uniform eccentricity distribution), with maximum initial eccentricity = 0.9, with otherwise identical input parameters as in Fig.~\ref{fig:circ}. The merger rate drops dramatically to $\mathcal{R}\sim 26.8\, {\rm Gpc}^{-3}{\rm yr}^{-1}$ as a delay time ($\sim 0.03$ Myr) is introduced since it takes time for gas damping to begin building up the circularized BH population. This population will also preferentially occur at thinner regions of the AGN disk, so the migration swamp/trap region remains important but fewer (less massive) IMBH are now produced.

\subsection{Testing dynamics}
Among a disk BH population where dynamics is ignored, BBH will tend to form in regions where there is crowding (e.g. near migration traps), as in Fig.~\ref{fig:ecc}. However, things can change once we allow dynamical interactions to occur.

First, we introduce interactions between the embedded BH in the disk. Circularized BH are dynamically heated by close encounters with the eccentric population. This introduces an additional delay time ($\sim 0.3$Myr, excluding a handful of very close initial BH) before BBH can merge in addition to the gas circularization delay time seen in Fig.~\ref{fig:ecc}. The average rate of dynamical encounters among an eccentric BH population is highest in the inner disk and so the circularized population build-up near the migration trap/swamp is inhibited initially, leading to a delay in higher generation mergers. This is clear from Fig.~\ref{fig:dyn_ecc}
 where we assume $e_{\rm max}=0.3$ and the BH populations within the disk (eccentric and circularized) are allowed to interact with each other. Dynamical destruction of inner disk BBH leads to a delay in 3g-ng mergers, which first appear after $\sim 0.4$Myr vs $\sim 0.2$Myr in Fig.\ref{fig:ecc}). As we increase $e_{\rm max}$, the delay-time increases and the AGN needs to run for longer to drive a higher merger rate. Fig.~\ref{fig:jm} shows the effect of choosing a different torque prescription but all else as in Fig.~\ref{fig:dyn_ecc}. The \citet{JimenezMassset17} prescription leads to a lower torque magnitude by $\sim 30\%$ across the outer disk, delaying mergers and inhibiting higher generation mergers for an AGN of the same lifetime (again see Fig.2 of \citep{Grishin23} for an approximate illustration).

\begin{figure} \includegraphics[width=1\columnwidth]{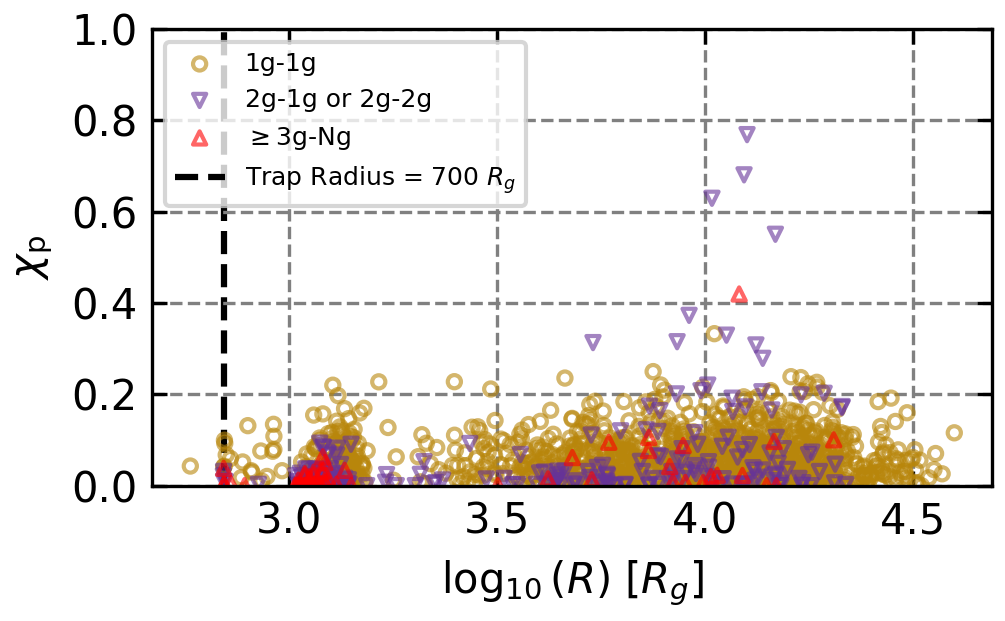}
      \includegraphics[width=1\columnwidth]{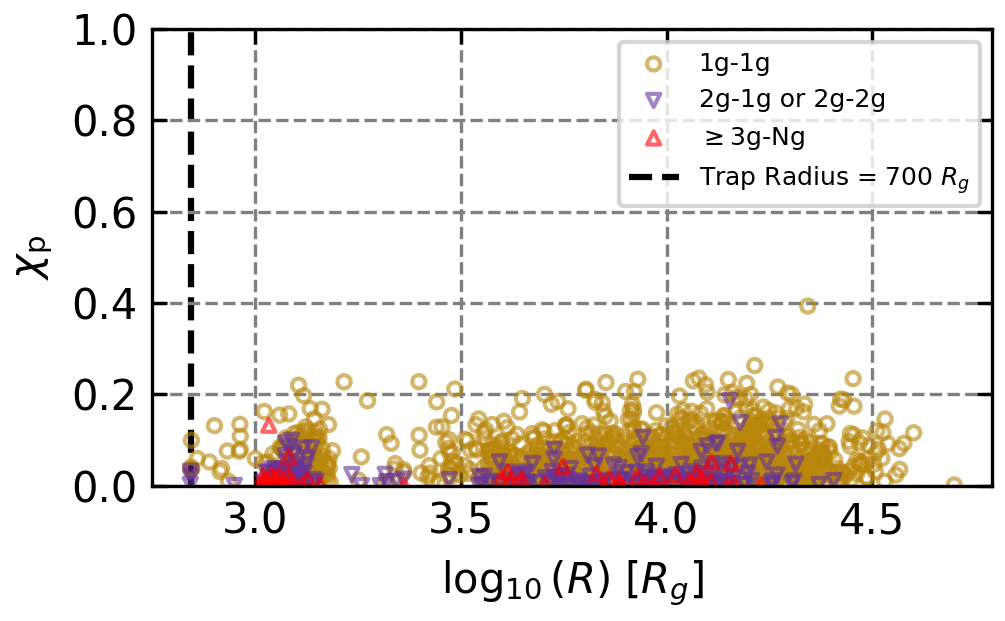}
    \caption{$\chi_{\rm p}$ for BBH mergers as a function of disk radius where spheroid encounters are on (top panel) and spheroid encounters are off (bottom panel). Input file and parameters (top panel) are as in Fig.~\ref{fig:dyn_ecc}, and in bottom panel except \texttt{nsc$\_$spheroid$\_$normalization = 0.0}. Equivalent merger rate is ${\mathcal{R}} \sim 22.0(22.7) \,{\rm Gpc}^{-3} {\rm yr}^{-1}$ top (bottom) panel.
   }
    \label{fig:chi_p}
\end{figure}


\begin{figure}   \includegraphics[width=1\columnwidth]{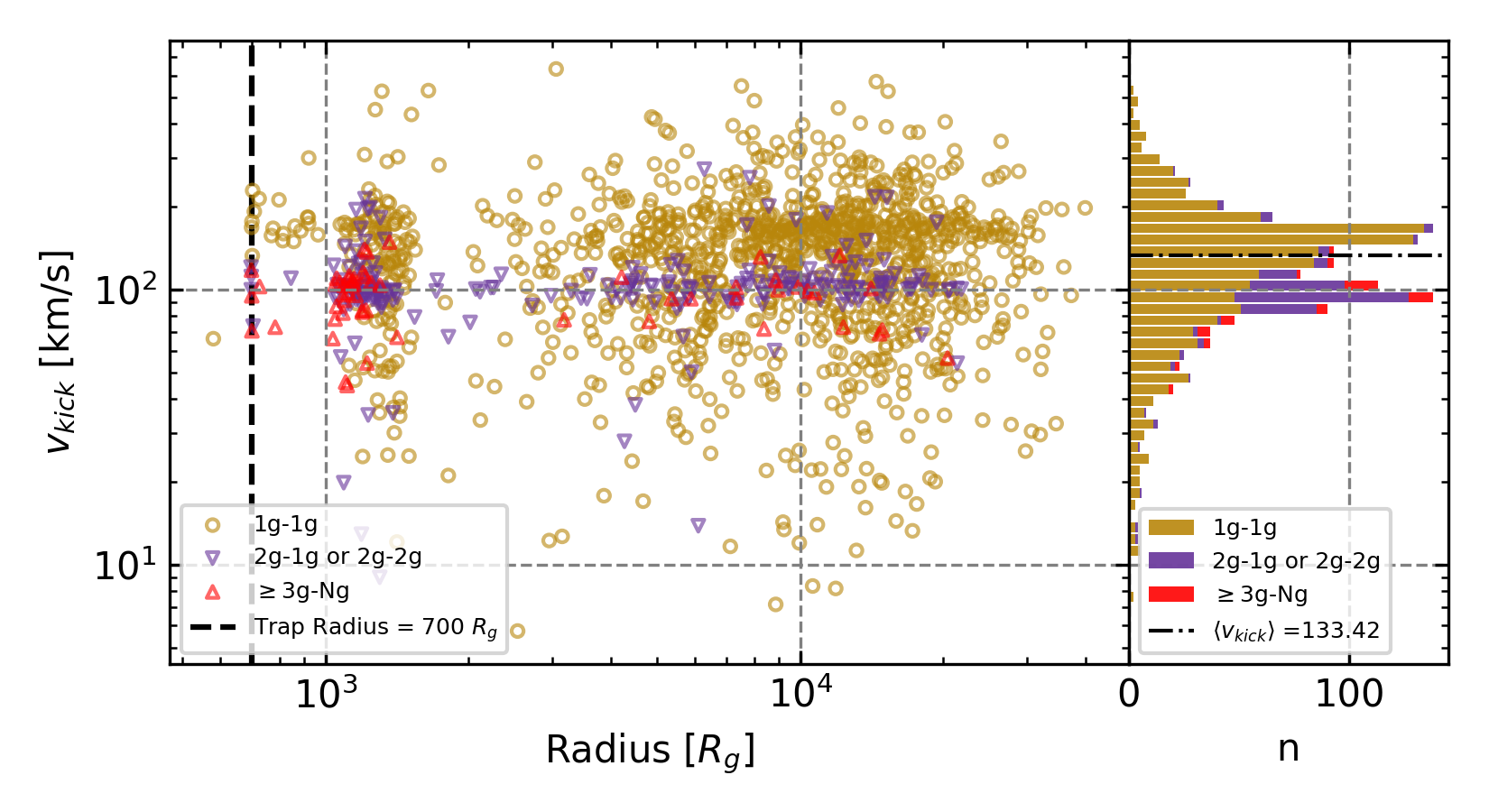}
    \caption{(\emph{Left panel:})  Velocity kicks as a function of disk radius and generation associated with mergers for the model in Fig.~\ref{fig:mf2}(a). (\emph{Right panel:}) Number and generation of mergers as a function of kick velocity. Dash-dot line corresponds to mean kick velocity of $\sim 133{\rm km/s}$.
    }
    \label{fig:v_kick}
\end{figure}
Dynamically we also introduce encounters between the spheroid NSC population and the embedded BBH population in the disk. Spheroid encounter rates are highest in the inner disk, but with disk dynamics on, the overall effect on the merger rate is small (few $\%$ change). 
Spheroid encounters have a significant effect in generating a in-plane spin component ($\chi_{\rm p}$). Conservation of orbital angular momentum in moderately close interactions between disk BBH and spheroid stars will pop BBH out of the AGN disk with a new orbital angular momentum ($L_{\rm BBH}$) orientation, but individual BH spin alignments do not change, yielding a significant spin component in the plane of $L_{\rm BBH}$ \citep{Hiromichi20spin,Samsing22,McKF24}. Fig.~\ref{fig:chi_p} shows $\chi_{\rm p}$ as a function of BBH merger location in the disk with spheroid encounters on/off (left/right panel respectively) (\texttt{nsc$\_$spheroid$\_$normalization=1/0}). The highest values of $\chi_{\rm p}$ are overwhelmingly driven by higher generation (higher mass) mergers. Higher mass BBH have a larger cross-section for spheroid encounters (larger $R_\mathrm{H}$), but higher mass makes them harder to ionize via spheroid stellar encounters. This feature of the AGN channel should be directly testable in O4 by LVK.

Fig.~\ref{fig:v_kick} shows the distribution of merger kick velocities as a function of disk radius, calculated from eqn.~\ref{eq:v_kick}. First generation BH mergers (in gold) peak strongly around $\sim 200{\rm km/s}$ as we expect \citep{Vijaykick}. However, higher generation BH have smaller kick magnitudes ($\sim 100{\rm km/s}$) with this prescription. We caution that this is an \emph{underestimate} of the kicks to be expected among higher generation mergers. In particular, including the relativistic evolution of the BBH allows for much higher values of the merger kick (up to $\sim 2000{\rm km/s}$) among the higher generation mergers. This will be very important for potential mass segregation between AGN phases \citep[e.g.][]{Gilbaum25} (see also Ray et al. 2025, in prep.).

\subsection{The ($q,\chi_{\rm eff}$) anti-correlation}
\label{sec:qXeff}
LVK observations suggest an intruiging anti-correlation between the BBH mass ratio ($q$) and the effective spin ($\chi_{\rm eff}$) of mergers \citep{Callister21}. The anti-correlation is consistent with greater alignment between spin and BBH orbital angular momentum for the primary BH component ($\chi_{1}$) when BBH masses are less equal. The only reason for such a anti-correlation to arise in a dynamical channel is if there is a bias towards alignment of the primary BH spin with the BBH orbital angular momentum. AGN disks can naturally provide such a bias \citep[e.g.][]{qX22, Santini23}. This bias alone (via spin-up and spin torquing) is insufficient however; a bias \emph{against} retrograde BBH mergers is also required. In the AGN channel, such a bias might occur because: (i) Retrograde disk BBH are flipped to prograde over time due to gas accretion \citep{AlexD24} or, (ii) retrograde BBH experience orbital eccentricity pumping \citep{Calcino23}, so retrograde BBH should spend more time at wider separations than prograde BBH with similar semi-major axes around their center of mass \citep{Yihan21}, or (iii) spheroid encounters with in-plane retrograde BBH will drive the BBH away from retrograde \citep{Samsing22} and gas torquing tends to drive the BBH towards $\chi_{\rm eff}>0$. 

\begin{figure*}
    \centering
    \includegraphics[width=1\columnwidth]{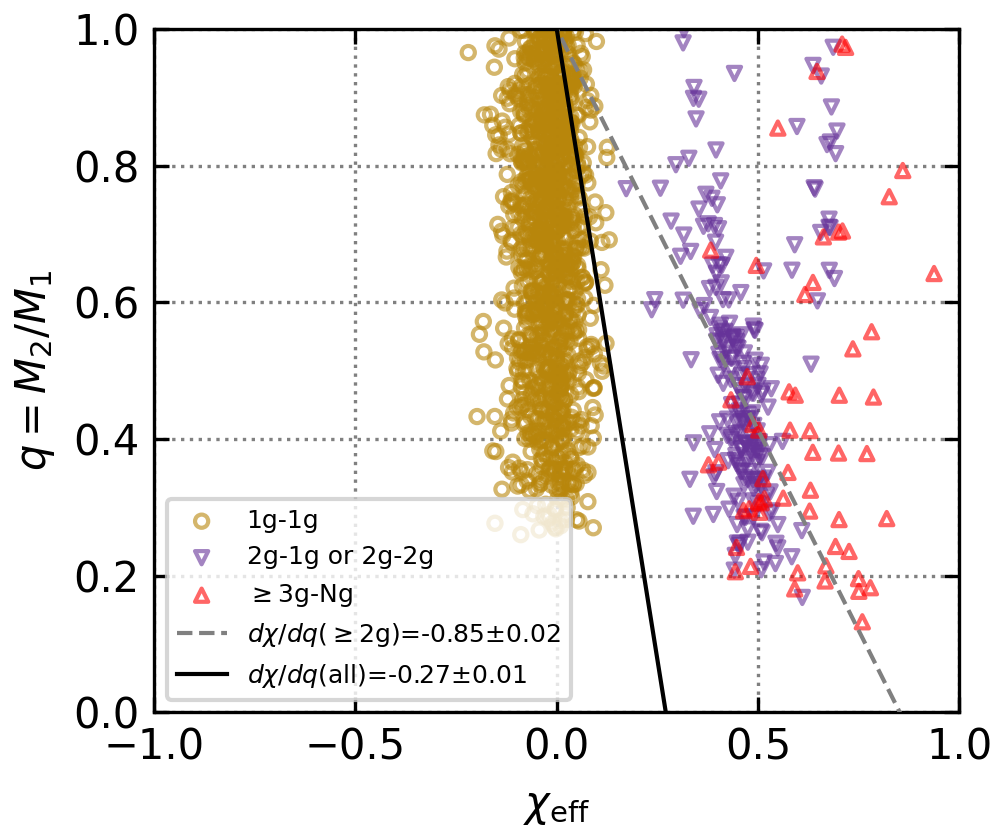}
      \includegraphics[width=1\columnwidth]{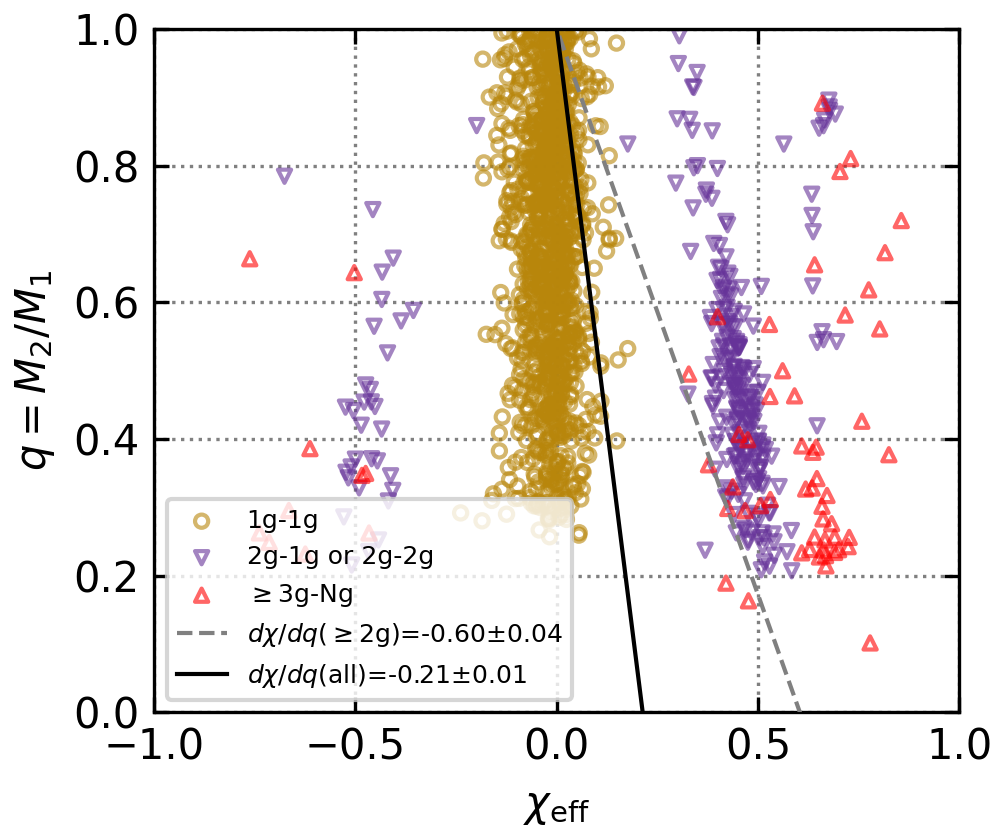}
    \caption{Mass ratio ($q$) distribution as a function of $\chi_{\rm eff}$. Left panel assumes all BBH are biased to form prograde  and right panel assumes $10\%$ form retrograde. Input file is \texttt{model$\_$choice$\_$old.ini} as for Fig.~\ref{fig:dyn_ecc} left panel and in the right panel we assume \texttt{fraction$\_$bin$\_$retro=0.1}.  Equivalent merger rates are ${\mathcal{R}} \sim 22.0(22.2) {\rm Gpc}^{-3} {\rm yr}^{-1}.$
   }
    \label{fig:qX}
\end{figure*}

\begin{figure*}
    \centering
    \includegraphics[width=1\columnwidth]{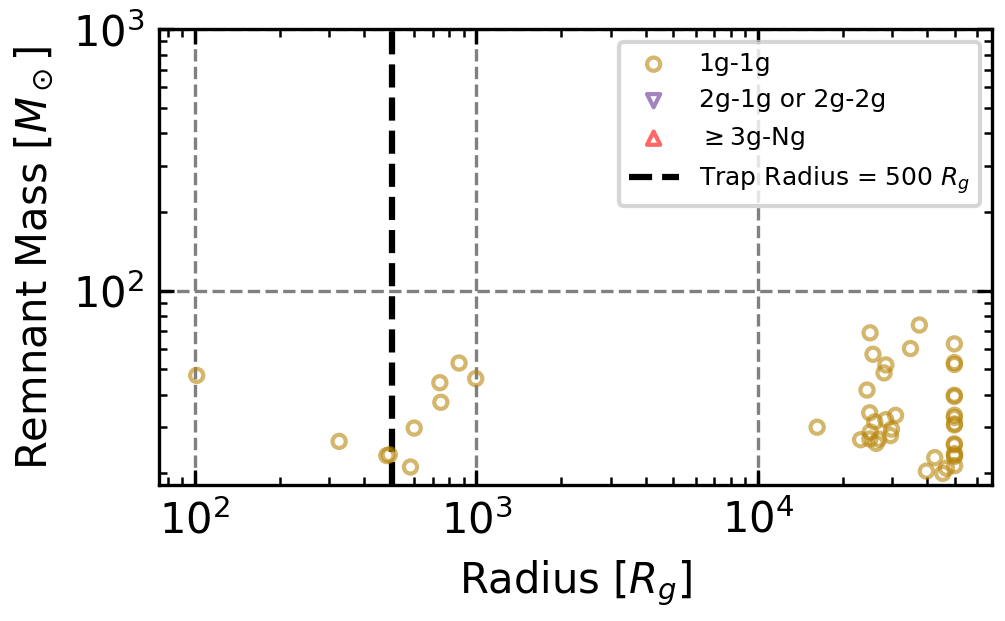}
      \includegraphics[width=1\columnwidth]{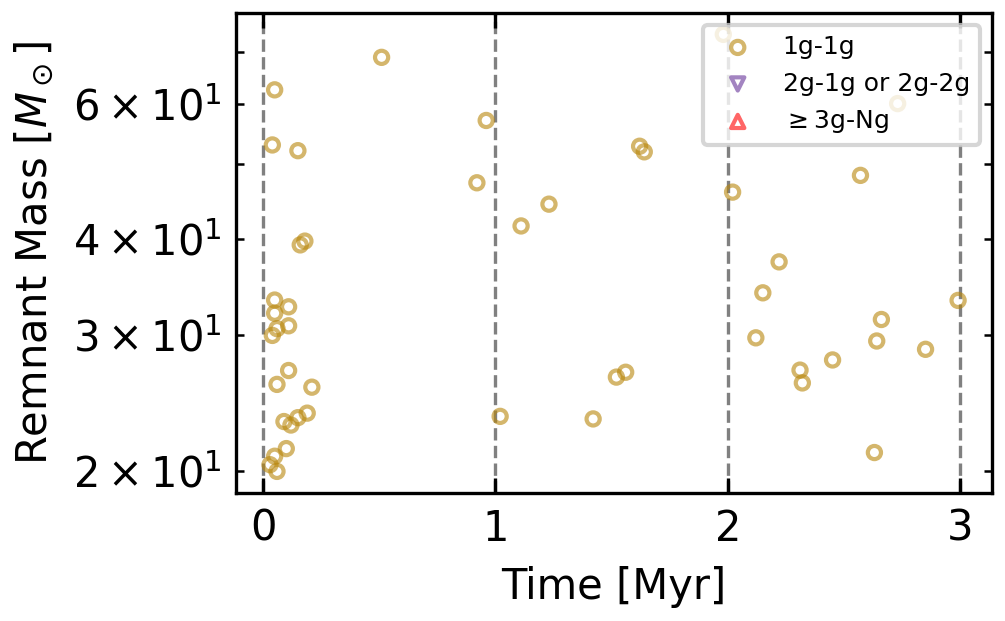}
    \caption{As Fig.~\ref{fig:dyn_ecc} except uses a \texttt{pAGN} supplied  \citet{TQM05} disk model for 3Myr. Input file is as in Fig.~\ref{fig:dyn_ecc} except \texttt{disk$\_$model$\_$name = 'thompson$\_$etal'}. Equivalent merger rate is ${\mathcal{R}} \sim 0.1  {\rm Gpc}^{-3} {\rm yr}^{-1}.$
   }
    \label{fig:tqm_default}
\end{figure*}

\begin{figure*}
    \centering
    \includegraphics[width=1\columnwidth]{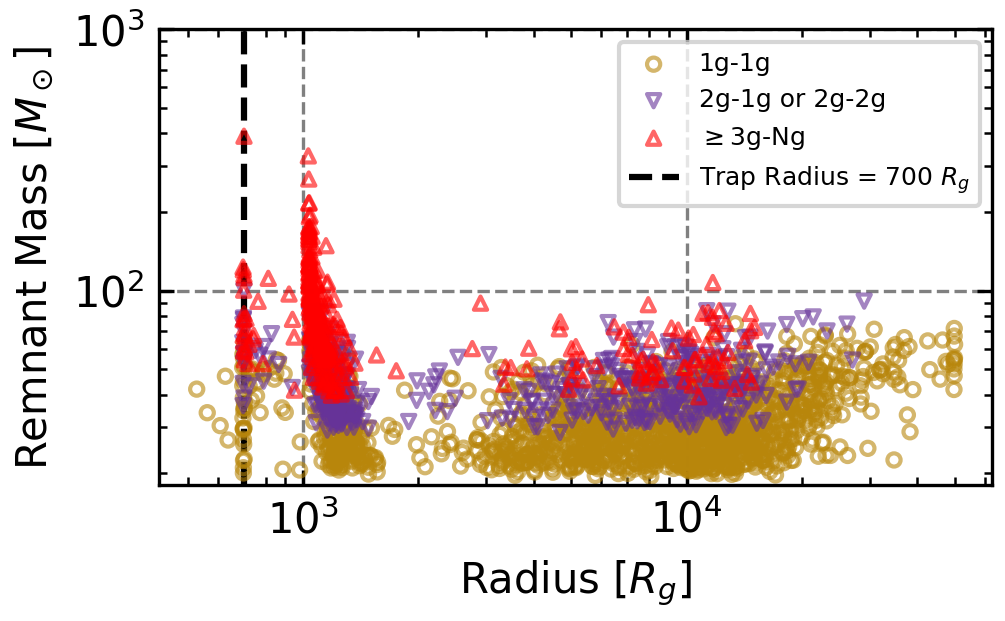}
      \includegraphics[width=1\columnwidth]{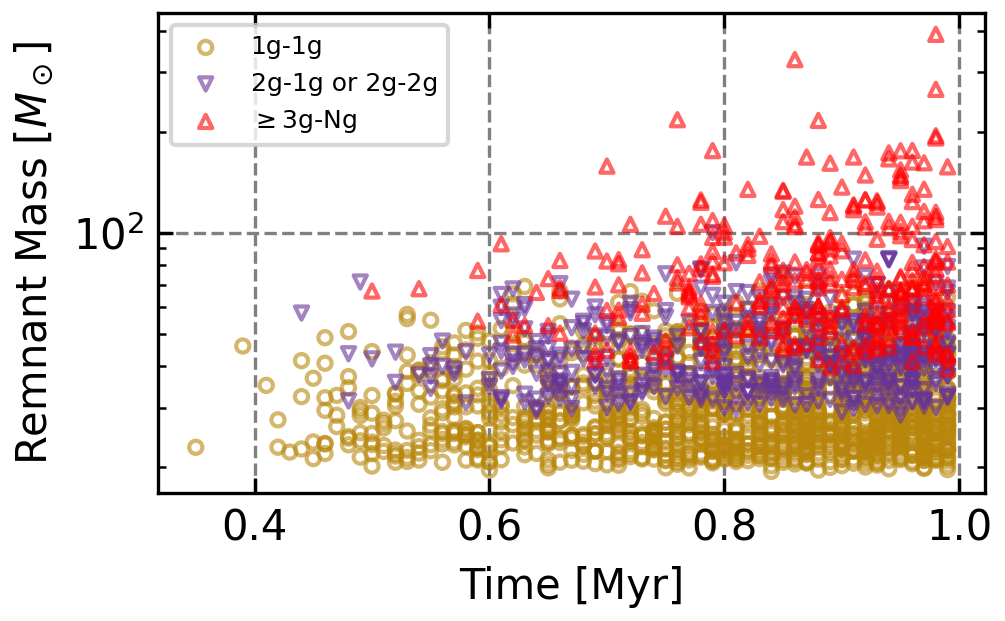}
    \caption{As Fig.~\ref{fig:dyn_ecc} except disk runs for 1Myr. Input file and parameters are as in Fig.~\ref{fig:dyn_ecc} except \texttt{timestep$\_$num = 100}. Equivalent merger rate is ${\mathcal{R}} \sim 41.2 {\rm Gpc}^{-3} {\rm yr}^{-1}.$
   }
    \label{fig:sg_dyn_on_5Myr_time}
\end{figure*}

Fig.~\ref{fig:qX} illustrates the ($q,\chi_{\rm eff}$) distribution for our default model, assuming the retrograde:prograde fraction of BBH mergers is $0(0.1)$ in the left(right) panel. Most mergers are 1g-1g and center around $\chi_{\rm eff}=0$. Higher generation mergers ($ng-mg$) ($n\geq 2,m \geq 1$) are mostly displaced to $|\chi_{\rm eff}| \sim [0.3,0.7]$ with more massive primaries typically having a spin that dominates the $\chi_{\rm eff}$ calculation. The width of the 1g-1g distribution depends on the choice of initial BH spin magnitude distribution and the extent of the 1g-1g distribution in $q$ is a function of the BH IMF ($M_{\rm BH,min/max}$). Thus, the ($q,\chi_{\rm eff}$) distribution is an effective test of the BH IMF and spin distribution in galactic nuclei. A full parameter space study can be found in companion paper, \citep{Cook24}.

\subsection{Testing AGN disk models}
\label{sec:disks}
We picked two AGN disk models to test, namely \citet{SG03} and \citet{TQM05}. Our default model is \citet{SG03} since this tends to give a better fit to the UV `big blue bump' in quasar spectral energy distributions and we expect that dense quasar disks are more efficient at coagulating embedded populations. In the case of \citet{TQM05}, this disk model is less dense, less optically thick and at least nominally allowed to have a migration trap at $500r_{g}$ \citep{Bellovary16}, although this is likely due to an incorrect modelling of opacities \citep{Alex20}.

Fig.~\ref{fig:tqm_default} shows the distribution of BBH masses as a function of disk radius (left panel) and merger time (right panel) for our default \citet{TQM05} model. Evidently, a less dense disk drives a significantly lower average merger rate ($\sim 0.1 \,{\rm Gpc}^{-3}{\rm yr}^{-1}$) than the \citet{SG03} model and the formation of new BBH becomes more difficult. The LVK AGN channel is likely filled with contributions from both high and low density disks.  Thus, in future work, we will introduce a prior spheroid binary fraction, which could be hardened towards merger, even in very low density AGN disks and which might dominate their contribution to the observed rate.

\subsection{Towards constraints on AGN lifetime}
\label{sec:time}
Fig.~\ref{fig:sg_dyn_on_5Myr_time} shows BBH merger masses as a function of location in disk and time for our default model run for 1Myr with $\rm{e}_{\rm max}=0.9$. The patterns observed in the shorter lived disk model Fig.~\ref{fig:dyn_ecc} persist (left panel) and mergers are delayed by $\sim 0.4{\rm Myr}$ by damping from a higher average initial eccentricity (right panel). The average equivalent merger rate is $\mathcal{R} \sim 41.2\,{\rm Gpc^{-3} yr^{-1}}$. As we limit the AGN fraction of LVK mergers, we will gain strong constraints on the average $\tau_{\rm AGN}$ and we will explore this in future work.

\subsection{Testing GW multi-wavelength models}
\label{sec:gw}
BBH that merge in AGN will be detectable across multiple GW frequencies ($\nu_{\rm GW}$) given by
\begin{equation}
    \nu_{\rm GW} = \frac{G^{1/2}M_{\rm BBH}^{1/2}}{\pi a_{\rm BBH}^{3/2}}
\end{equation}
where $G$ is the universal gravitational constant, $M_{\rm BBH}$ is the BBH total mass and $a_{\rm BBH}$ is the semi-major axis of the BBH around its own center of mass. The resulting gravitational wave strain is
\begin{equation}
    h = \sqrt{\frac{32}{5}} \frac{G^{2}}{c^{4}}\frac{M_{\rm BBH}\mu_{\rm BBH}}{D a_{\rm BBH}}
\end{equation}
with $c$ the speed of light, $\mu_{\rm BBH}=M_{1}M_{2}/M_{\rm BBH}$ is the BBH reduced mass, where $M_{1,2}$ are the masses of the individual BBH components and $D$ is the distance to the BBH. Importantly, in AGN, binary black holes (BBH) can form between two stellar mass BH, but also between individual stellar mass BH and the central SMBH. The latter (SMBH-BH binaries) will be detectable at low GW frequencies as extreme mass ratio inspirals (EMRIs) with LISA \citep{LISA23}.

As BBH evolve (due to gas hardening/softening, binary softening/hardening or GW emission) they change their GW frequency. A useful parameter, the characteristic strain, averaged over all polarizations is
given by $h_{\rm char} = (N/8)h$ where $N = \sqrt{2\nu_{\rm GW}^{2}/\dot{\nu}_{\rm GW}}$ and $\dot{\nu}_{\rm GW}=\mathcal{M}_{\rm ch}^{5/3}\nu_{\rm GW}^{11/3}$ where the binary chirp mass $\mathcal{M}_{\rm ch}=(M_{1}M_{2})^{3/5}/M_{\rm BBH}^{1/5}$ if $\nu_{\rm GW}>10^{-6}\,{\rm Hz}$. 

\begin{figure}   \includegraphics[width=1\columnwidth]{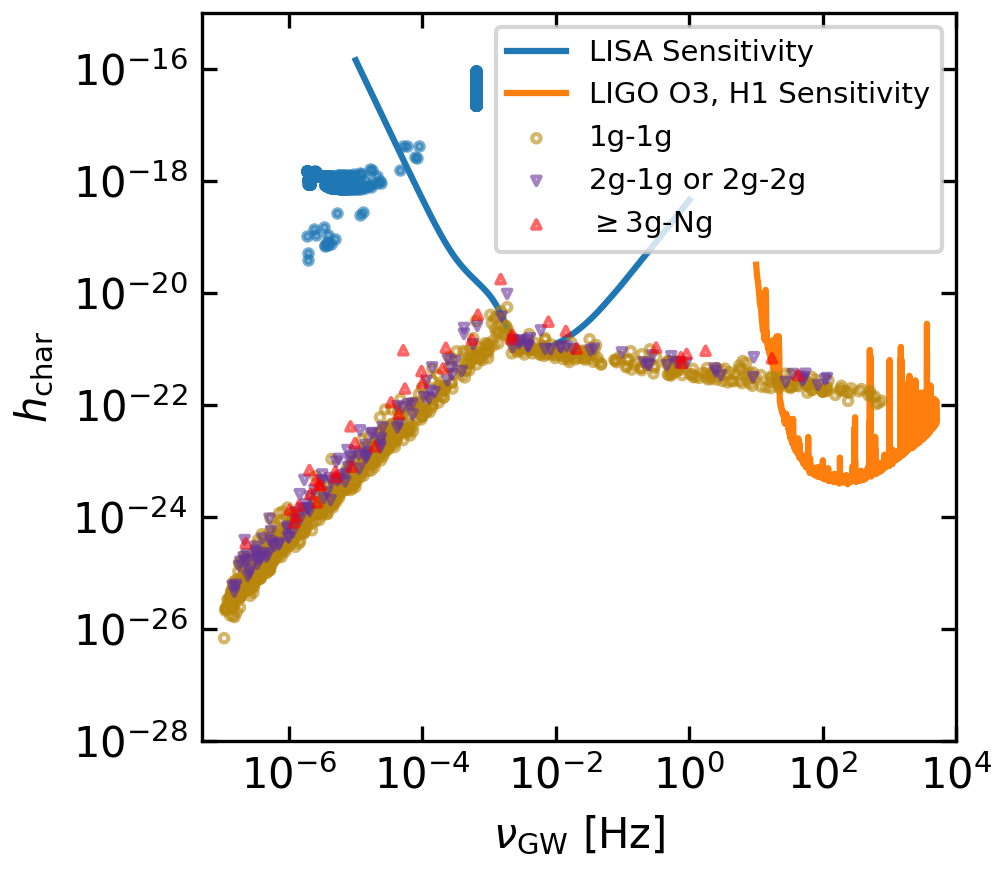}
    \caption{Characteristic strain ($h_{\rm char}$) as a function of GW frequency ($\nu_{\rm GW}$) after a 0.5Myr episode with default AGN model assumed to lie at $z=0.1$. Note that the LVK sensitivity curve is appropriate for $h_{0}$ rather than $h_{\rm char}$ (see Postiglione et al. 2025 in prep. for a more accurate treatment). Input file and parameters are as in Fig.~\ref{fig:dyn_ecc}.
    }
    \label{fig:gw_strain}
\end{figure}

Fig.~\ref{fig:gw_strain} shows the characteristic strain  ($h_{\rm char}$) as a function of gravitational wave frequency ($\nu_{\rm GW}$) assuming all AGN are located at $z=0.1$ by default. For detailed simulations and predictions of redshift dependency of the outputs produced by \texttt{McFACTS} as a function of $M_{\rm SMBH}$ and NSC properties see \citep{Delfavero24}. Curves in blue (orange) correspond to the LISA (LIGO Hanford O3) sensitivity curves respectively. Note that the LIGO curve is appropriate for $h_{0}$ rather than $h_{\rm char}$. Points in blue correspond to snapshots at the end of each default timestep of EMRIs from 100 iterations of the default model (see Ford et al. 2025 in prep.). Points in gold, purple, red correspond to snapshots of 1g,2g and $\geq 3$g BBH at the end of each timestep of individual BBH with $\nu_{\rm GW}>10^{-7}\,{\rm Hz}$. 

On average $2-3$ points are associated with each BBH, since our default gas hardening prescription \citep{Baruteau11} drives rapid mergers of hard binaries. Once a BBH shrinks to a point where it will merge within the next timestep ($\mathcal{O}(10^{-3}{\rm Hz})$ for default), $\dot{\nu}_{\rm GW}$ in $N$ above is switched from gas-hardening to GW hardening \citep{Peters64}, leading to the break in $h_{\rm char}$ around $\sim 10^{-3}{\rm Hz}$  The first BBH point is arrival in the LISA frequency band on a single timestep, a second BBH point is the evolved BBH in the next timestep, which could correspond to a merger in the LIGO band, or a dynamically hardened/softened BBH (see Postiglione et al. 2025 for a detailed study, and for an illustration of the range of possible BBH merger tracks). BBH that enter the LISA band, but are then ionized, are typically represented by a single point on this plot.  The band of gold, red and purple points correspond to snapshots of BBH mostly gas-hardening towards merger with typical value of $h_{\rm char} \approx \mathcal{O}(10^{-22})$. 
Since binary hardening or softening in the code is either gas-driven or dynamics driven, we have not yet combined gas and dynamics hardening in a given timestep. More sophisticated treatment of both gas and dynamical hardening will be treated in future work ({Postiglione et al. 2025 in prep.}), particularly with a view to testing LISA detectability of gas and dynamically hardened AGN channel BBH. 

In \texttt{McFACTS}, EMRIs form when a BH arrives within $a<50\,r_{g}(M_{\rm SMBH}/10^{8}\,M_{\odot})^{-1}$ of the SMBH. The rate at which BH appear at such small disk radii is strongly effected by two factors: first, the number of retrograde embedded BH initially in the disk and second, the rate of disk capture and the range of allowed capture radii. 

When an AGN disk arrives in a relaxed galactic nucleus population, around half of the initial population of orbiters embedded in the disk will have orbits that are retrograde with respect to the disk gas. Depending on their semi-major axis and the disk surface density profile, the gas can excite the orbital eccentricity of embedded retrograde BH to large values, potentially generating a population of BH at small disk radii \citep{Secunda21}. However, perturbations of orbital inclination of embedded retrograde orbiters, due to e.g. turbulence, or interactions with spheroid orbiters, will tend to drive the retrograde orbiters out of the disk and tend to rapidly flip them to prograde \citep{WZL24}. This initially retrograde population in a turning-on AGN will tend to drive a high EMRI rate (these correspond to the vertical line of blue points at $\nu_{\rm GW} \sim 10^{-3}{\rm Hz}$ in Fig.~\ref{fig:gw_strain} and see {Ford et al. 2025 in prep.}). After multiple short-lived AGN episodes in the same nucleus, the retrograde fraction in the disk should become very small, so the LISA EMRI rate should be highest in new AGN episodes in relatively relaxed galactic nuclei (Ford et al. 2025 in prep.).

Under default models, the disk captures BH uniformly at radii $a<2\times 10^{3}\,r_{g}$ every 0.1Myr, so there is a $2.5\%$ chance of an EMRI every disk capture under default assumptions. EMRIs are considered to only evolve under the evolution of GW emission (gas effects are ignored for now). So in Fig.~\ref{fig:gw_strain} the characteristic strain per frequency (equivalently the amount of time at a monochromatic 
 GW frequency) decreases as the EMRI shrinks (as $\nu_{\rm GW}$ increases). The exception occurs during the chirp at merger where $h$ increases dramatically. Some EMRIs form late in a run or at large enough distance ($\sim 50\,r_{g}$) and do not chirp within the 0.5Myr lifetime. These correspond to the population at lower characteristic strain ($h_{\rm char}/\nu_{\rm GW} \sim 10^{-19}$ in Fig.~\ref{fig:gw_strain}).

\section{Future Work}
\label{sec:future}
\texttt{McFACTS} is designed to be public, modular and evolve. Future near-term goals include: (i) add objects other than BH to the disk (neutron stars, white dwarfs, and stars) (see Nathaniel et al. 2025 in prep.) (ii) consider the EM significance of encounters between objects (including TDEs, $\mu$ TDEs, partial TDEs, GRBs as well as disk-crossing events), (iii) treat the final timestep of mergers with increasing sophistication, including calculating details of eccentricity evolution, recoil kicks from numerical relativity calculations (Ray et al. 2025 in prep.), and disk re-capture, (iv) allow multiple AGN episodes to occur in a given nucleus, with appropriate dynamical relaxation, (v) predict the rate and magnitude of optical/UV/X-ray flaring due to explosive events (mergers, disk-crossing, SNe, TDEs) (McPike et al. 2025 in prep.), (vi) develop and add new models of AGN disks to reflect recent simulations \citep{FIRE24} and (vii) add more sophisticated treatments for binary capture \citep{Whitehead25} and stalling of gas hardening (Postiglione et al. 2025 in prep.). We also welcome suggestions and contributions from the community for code additions and improvements as we continue development.

\section{Conclusions}
We demonstrate the new, public, open source, fast and reproducible code $\texttt{McFACTS}$ , the Monte Carlo For AGN channel Testing and Simulation. \texttt{McFACTS} allows rapid testing of BBH merger properties across AGN and NSC parameter space. Users can choose their own AGN disk model or NSC model and quickly get population statistics and distributions of parameters. The code as released (\texttt{v.0.3.0}) makes multiple simplifying assumptions but is intended to be a living, developing code (see \S\ref{sec:future} for planned near-future additions).

We illustrate the code and demonstrate its performance by varying AGN and NSC parameter inputs and show that a high average rate ($\mathcal{R}$) of BBH mergers in the AGN channel is generally associated with relatively dynamically cool populations of BBH in short-lived AGN. This may imply a strong contribution to the LVK AGN channel from AGN fuelled by pulses of low angular momentum gas in the same plane from a fuel reservoir. For detailed studies of the AGN channel in ($q,\chi_{\rm eff}$) parameter space and as a function of NSC model, Galaxy mass and redshift, see companion papers \citep{Cook24}and \citep{Delfavero24} respectively. 
\begin{acknowledgments}
Thanks to the referee for an excellent report that helped improve the code and paper. BM, KESF and HEC are supported by NSF AST-2206096. BM \& KESF are supported by NSF AST-1831415 and Simons Foundation Grant 533845 as well as Simons Foundation sabbatical support and support to release this code. The Flatiron Institute is supported by the Simons Foundation. 
ROS gratefully acknowledges support from NSF awards NSF PHY-1912632, PHY-2012057, PHY-2309172, AST-2206321, and the Simons Foundation.
KN thanks the LSST-DA Data Science Fellowship Program, which is funded by LSST-DA, the Brinson Foundation, and the Moore Foundation; her participation in the program has benefited this work.
VD is supported by an appointment to the NASA Postdoctoral Program at the NASA Goddard Space Flight Center administered by Oak Ridge Associated Universities under contract NPP-GSFC-NOV21-0031.
\end{acknowledgments}

%

\vspace{5mm}


\software{Astropy, \citep{2013A&A...558A..33A,2018AJ....156..123A}, pAGN \citep{pAGN24}, NumPy \citep{harris2020array}, SciPy \citep{2020SciPy-NMeth}, Matplotlib \citep{Hunter_2007}.
          }

\appendix
\section{Reproducibility}
Each figure produced in this paper can be reproduced using the correct seed and .ini file and the \texttt{make plots} command. In the \texttt{Makefile} under Setup, change the \texttt{SEED} parameter to the value for the figure and change the \texttt{FNAME$\_$INI} file to the required .ini file and any variable as listed in the Figure caption.





\bibliography{refs}{}
\bibliographystyle{aasjournal}



\end{document}